\documentclass[draft]{agujournal}
\draftfalse

\journalname{<Geochemistry, Geophysics, Geosystems>}

\usepackage{wasysym}
\usepackage{hyperref}
\usepackage{longtable}
\usepackage{caption}

\begin{document}

%% ------------------------------------------------------------------------ %%
%
%  TITLE
%
%% ------------------------------------------------------------------------ %%

\title{EarthN: A new Earth System Nitrogen Model}

%% ------------------------------------------------------------------------ %%
%
%  AUTHORS AND AFFILIATIONS
%
%% ------------------------------------------------------------------------ %%

 \authors{Benjamin W. Johnson\affil{1,2} 
  Colin Goldblatt\affil{2}}

\affiliation{1}{Department of Geological Sciences, University of Colorado, 
Boulder, Colorado, USA }
\affiliation{2}{School of Earth and Ocean Sciences, University of Victoria, 
Victoria, British Columbia, Canada}

\correspondingauthor{B. W. Johnson}{benjamin.w.johnson@colorado.edu}

\begin{keypoints}
\item We model the evolution of nitrogen in all the reservoirs of Earth
\item Total, non-core N and plate tectonics exert strong control on atmospheric mass
\item Weathering and the Great Oxidation event cause atmospheric draw-down
\end{keypoints}

%% ------------------------------------------------------------------------ %%
%
%  ABSTRACT
%
%% ------------------------------------------------------------------------ %%

\begin{abstract}
The amount of nitrogen in the atmosphere, oceans, crust, and mantle have important ramifications for Earth's biologic and geologic history. Despite this importance, the history and cycling of nitrogen in the Earth system is poorly constrained over time. For example, various models and proxies contrastingly support atmospheric mass stasis, net outgassing, or net ingassing over time.  In addition, the amount available to and processing of nitrogen by organisms is intricately linked with and provides feedbacks on oxygen and nutrient cycles. To investigate the Earth system nitrogen cycle over geologic history, we have constructed a new nitrogen cycle model: EarthN. This model is driven by mantle cooling, links biologic nitrogen cycling to phosphate and oxygen, and incorporates geologic and biologic fluxes. Model output  is consistent with large (2-4x) changes in atmospheric mass over time, typically indicating atmospheric drawdown and nitrogen sequestration into the mantle and continental crust. Critical controls on nitrogen distribution include mantle cooling history, weathering, and the total Bulk Silicate Earth+atmosphere nitrogen budget. Linking the nitrogen cycle to phosphorous and oxygen levels, instead of carbon as has been previously done, provides new and more dynamic insight into the history of nitrogen on the planet.  
\end{abstract}

%% ------------------------------------------------------------------------ %%
%
%  TEXT
%
%% ------------------------------------------------------------------------ %%

\section{Introduction}
Despite its importance and abundance in the Earth system, relatively little is known about the cycling of N throughout the major reservoirs of the Earth through time \citep{Zerkle_and_Mikhail_2017}. This is an important component of the Earth system, as the amount of N in the atmosphere can directly affect the climate \citep{Goldblatt_et_al_2009, Wordsworth_and_Pierrehumbert_2013} as well as biologic productivity \citep{Klinger_et_al_1989}. Recent work has challenged the notion that N is primarily an atmospheric species, and instead the solid Earth may actually hold the majority of the planet's N budget \citep{Marty_2012, Halliday_2013, Johnson_and_Goldblatt_2015, Barry_and_Hilton_2016, Mallik_et_al_2018}. 

While the major biologic and geologic fluxes affecting N distribution are known, their behavior over Earth history is not constrained. Early descriptions of atmospheric  N$_2$ in the Precambrian admitted lack of data prevented speculation on what the atmospheric, and therefore mantle and continental crust, N content was at that time and how it has evolved since \citep{Delwiche_1977}.  Subsequent work generally supports three hypotheses: steady-state atmospheric N mass over time \citep{Marty_et_al_2013},  net mantle outgassing over time \citep{Som_et_al_2012, Som_et_al_2016}, and net ingassing over time \citep{Nishizawa_et_al_2007,Goldblatt_et_al_2009, Johnson_and_Goldblatt_2015, Barry_and_Hilton_2016, Mallik_et_al_2018, Yoshioka_et_al_2018}. Importantly, the assumption that atmospheric mass should be constant over Earth history is not an inherent property of the planet. 

Preliminary modelling efforts  considered sedimentary rocks as the main geologic storage and recycling vector for N \citep{Zhang_and_Zindler_1993, Berner_2006} and compared N geochemically to the noble gases or carbon \citep{Tolstikhin_and_Marty_1998}. These studies found that there was little change ($<1\%$) in atmospheric N$_2$ over at least the Phanerozoic \citep{Berner_2006} and possibly the majority of Earth history \citep{Zhang_and_Zindler_1993}. Additionally, while comparison to noble gases is valid for outgassing of oxidized magmas \citep{Libourel_et_al_2003}, this comparison is not valid at subduction boundaries, as N is mostly found as NH$^+_4$in subducted sediments \citep{Bebout_and_Fogel_1992} and oceanic crust \citep{Busigny_et_al_2011}. 

\begin{figure*}
\includegraphics[keepaspectratio=true, width=\textwidth]{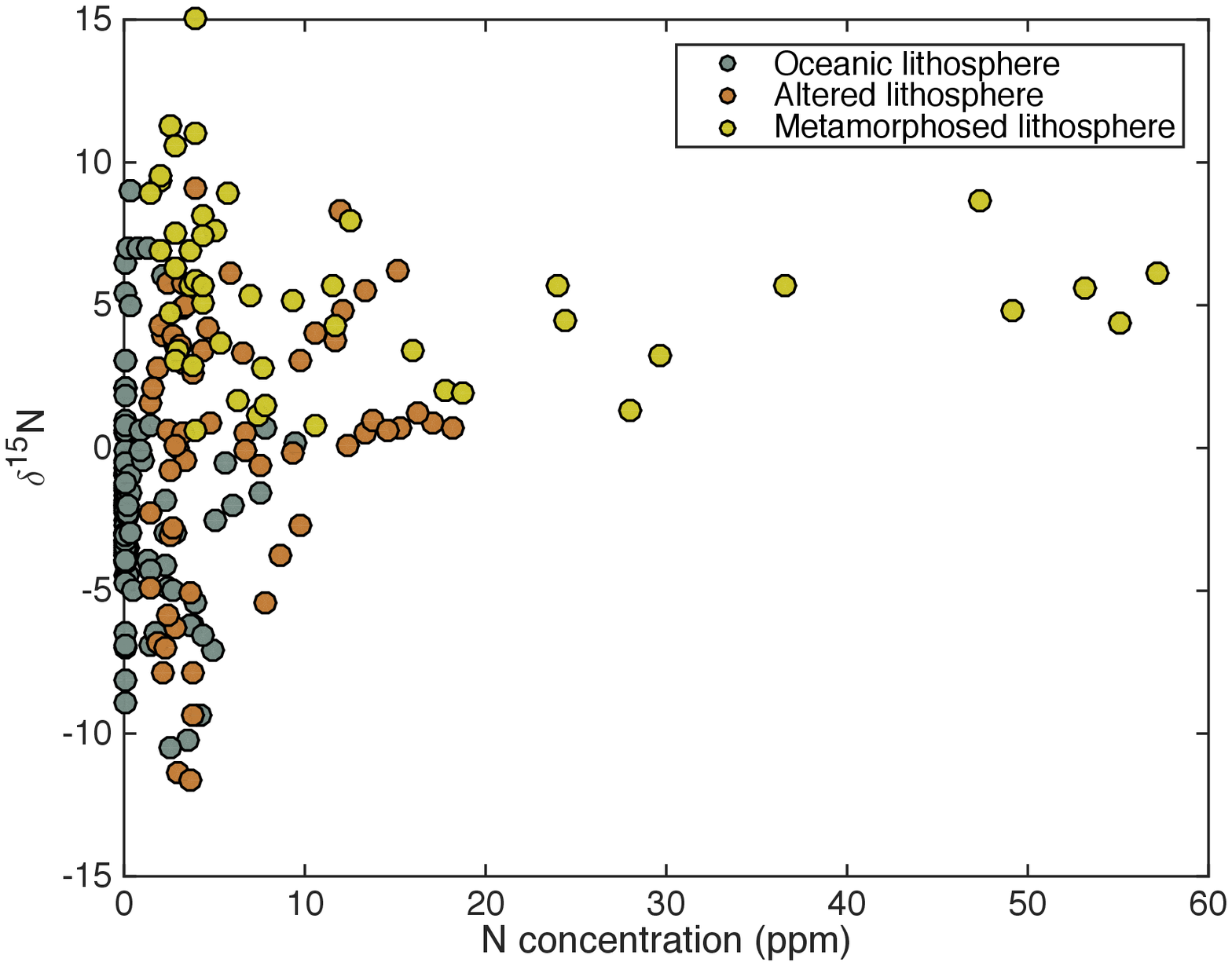}
\caption{$\delta^{15}$N plotted against N concentration (ppm) for oceanic lithosphere. Data are from the compilation of \citet{Johnson_and_Goldblatt_2015}. Shown are values from fresh lithosphere, altered lithosphere, and metamorphosed. Altered lithosphere has experienced temperatures of $<$300 $^{\circ}$C and metamorphosed samples have experienced temperatures $>$300 $^{\circ}$C. As oceanic lithosphere experiences alteration, N concentration increases, and initially depleted $\delta^{15}$N values, with a mean of -1.8\permil~ are enriched, indicating additional N is sourced from biologic material, which has $\delta^{15}$N of $5\permil$. Altered lithosphere rocks include MORBs while metamorphosed lithosphere rocks include blueschists, eclogites, and metagabbros.}
\label{fig:del15NMORB}
\end{figure*}

The geologic treatment of N in previous models may have missed some important behavior. Specifically, only considering sedimentary rocks as a sink for biologically processed N \citep{Berner_2006} based on N/C ratios does not include hydrothermal addition of N to oceanic crust (Fig. \ref{fig:del15NMORB}), which is observed in modern and older altered crust \citep[e.g.,][]{Halama_et_al_2014}. Sediment-only geologic N models also assume N and C behave similarly in subduction zones, which may not be true as N is likely found primarily as NH$^+_4$~ geologically and C as organic C or CO$_3^{2-}$. In addition, previous whole-Earth modeling \citep{Tolstikhin_and_Marty_1998} maintained a steady-state upper mantle, in terms of N-isotopes and concentration, by recycling of  sedimentary and sea-water-sourced N from the surface and N from the lower mantle entrained in plumes. This approach is incomplete, as, again, surface N is subducted to the mantle as NH$^+_4$~ and it is likely that the mantle as a whole is not layered. Therefore, a mechanism of ``re-filling'' the upper mantle from the lower mantle slowly over time appears untenable. 

Studies of several modern subduction zones suggest more dynamic N evolution, and that there is overall net transport of N into the solid Earth, either the mantle or arc-generated crust. Importantly, the N that survives the subduction barrier seems to mostly reside in the oceanic crust \citep{Li_et_al_2007, Mitchell_et_al_2010}. There are many possible mineral hosts for such N, typically found as NH$^+_4$, during subduction, including NH$^+_4$-bearing feldspars, pyroxenes, beryls, and phlogopite in the mantle \citep{Watenphul_et_al_2009, Watenphul_et_al_2010, Bebout_et_al_2015} . Such crystalline N in altered crust appears to be more likely to be carried into the mantle, whereas sedimentary N tends to return to the atmosphere at subduction zones \citep{Fischer_et_al_2002, Elkins_et_al_2006, Halama_et_al_2014}. 
 
 As such, we are presented with a conundrum. Modeling efforts suggest that the atmosphere and solid Earth have remained in equilibrium in terms of N-content over time. Contrastingly, geochemical evidence suggests there may be net transport of N from the surface to the mantle over time. It is from this conundrum that the construction of an Earth-system N cycle model, EarthN, follows. 

Previous Earth system models implicitly have biologic processing \citep[e.g.,][]{Stueken_et_al_2016}, but none so far actually explicitly model the behavior of organisms. Biologic productivity and activity is the gate-keeper between the atmosphere and the solid Earth.  Similarly, recent work has modeled the nitrogen cycle but without biology \citep{Laneuville_et_al_2018}, to serve as a background for interpretations and models including biology. Nitrogen can cycle throughout  the atmosphere, biosphere, sedimentary rocks, and crystalline Earth, thus constructing a model that integrates both biologic and geologic fluxes is critical for investigating the N-cycle over Earth history. 

 \begin{figure*}
 
\includegraphics[]{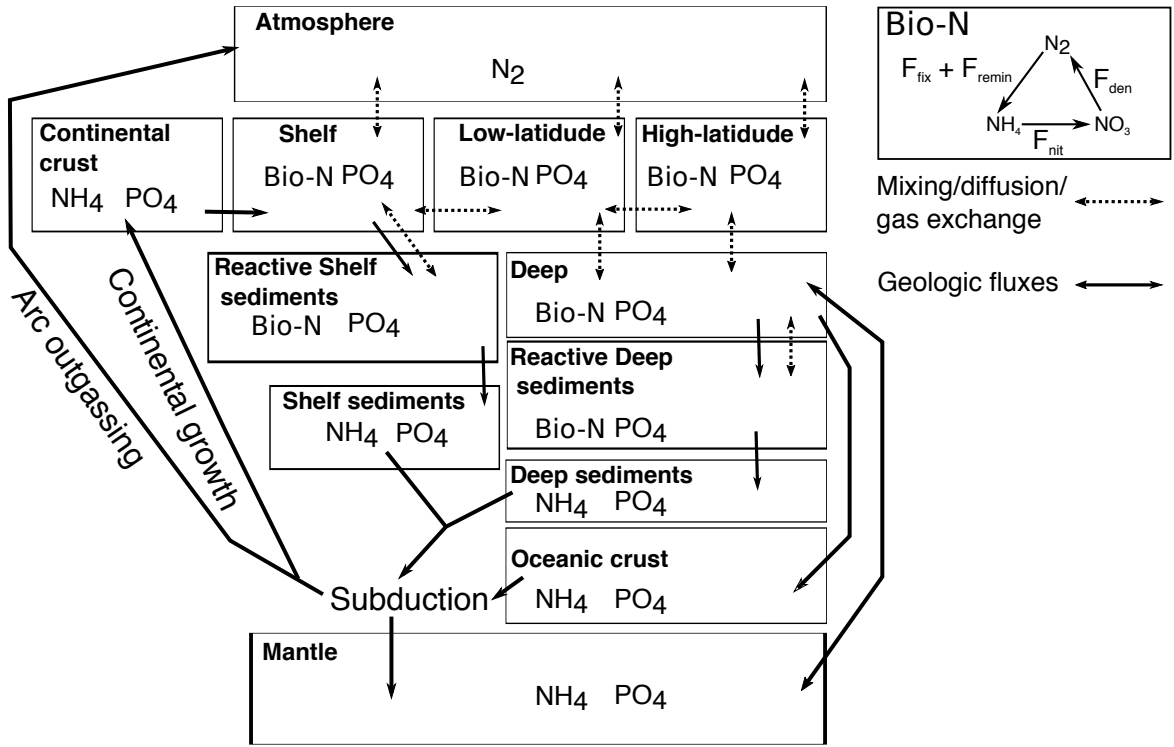}
\caption[Earth system nitrogen cycle model schematic]{Earth system nitrogen cycle model schematic. In addition, we included K, $^{40}$K, $^{40}$Ar, $^{36}$Ar, and $^{40}$Ca as checks on non-biologic cycles, though these are not shown here. }
\label{fig:model_schematic}
\end{figure*}

\section{Model setup}
The model is divided into a number of boxes. These are the atmosphere, three shallow ocean boxes (low-latitude, high-latitude, shelf), deep ocean, two biologically active sediment boxes (reactive shelf, reactive deep), sediments not in communication with the ocean (shelf, deep), and geologic reservoirs (mantle, oceanic crust, continental crust).

The model contains  N as N$_2$, NO$^-_3$, and NH$^+_4$, the last of which can be in the ocean or in geologic reservoirs. We also include other biologically relevant species: PO$^{3-}_4$ and O$_2$, as well as inorganic tracers: K,$^{40}$K,$^{40}$Ar, and$^{36}$Ar. Phosphate directly affects biologic productivity and O$_2$  affects both productivity and which pathways of the biologic N cycle are in operation. Nitrogen is geochemically similar to K when found as NH$^+_4$and geochemically similar to Ar when found as N$_2$. As K and Ar are not biologically important elements, they serve as both a calibration and validation of the purely physical aspects of the model (Appendix \ref{apdx:Ar_K}).

 \begin{figure*}
\includegraphics[keepaspectratio=true, width=\textwidth]{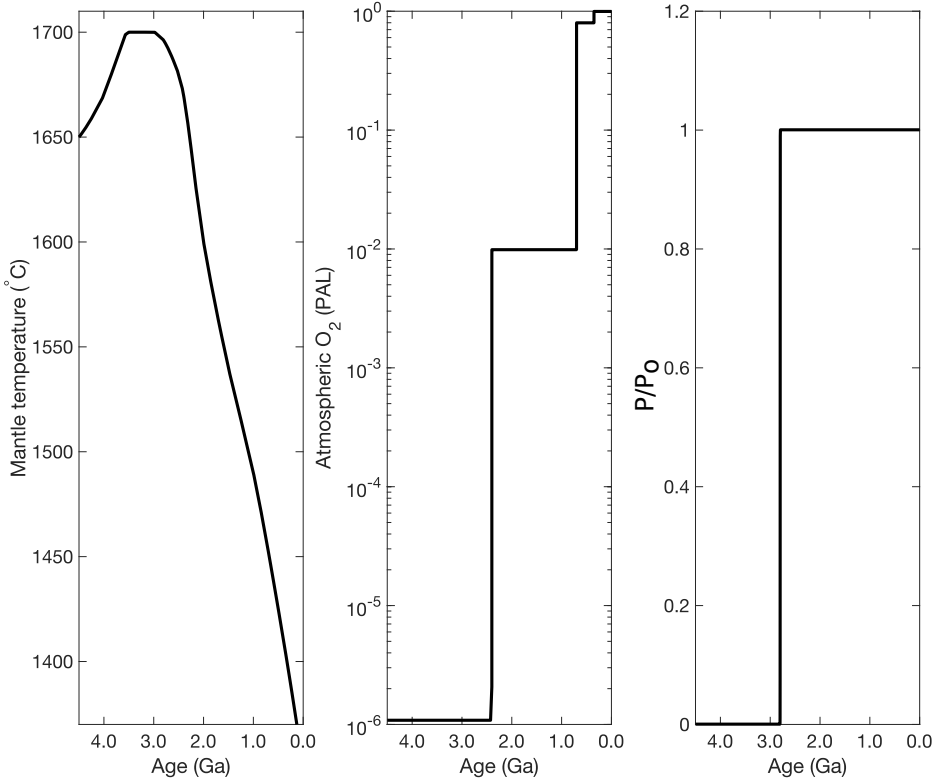}
\caption[Model forcings]{Model forcings, including average mantle potential temperature ($^{\circ}$C) \citep{Korenaga_2010}, atmospheric O$_2$ \citep[e.g.,][]{Lyons_et_al_2014}, and primary productivity (P) compared to modern (P$_\text{o}$) constant \citep{Goldblatt_et_al_2006}.}
\label{fig:model_forcings}
\end{figure*}

Biogeochemical fluxes are after \citet{Fennel_et_al_2005}, with a number of updates. Most geologic fluxes are driven by mantle cooling history after \citet{Korenaga_2010} and \citet{Padhi_et_al_2012}, which produces estimates of mantle temperature, crust production, and spreading rate through time (Fig. \ref{fig:model_forcings}). Some runs have a constant proportion of subducted material retained to the mantle, and some link subducted fraction to mantle temperature. 

The following is first a brief description of element cycles, then a detailed description of the model setup. We discuss each model box, the species contained within said box, and the fluxes that affect the amount of each species in the box. We use $R_j^i$ to represent reservoir  in moles size of species $j$ in box $i$ and  $C_j^i$ to represent the corresponding concentration. $F_k^i$  represents  fluxes  of type $k$ in box $i$ (Table \ref{tab:fluxes}). There are a number of fluxes that are sensitive to reactant concentration (e.g., Michaelis-Menten behavior), and are shown as $v_l$ where $l$ is specific to each sensitivity $v$. Parameter values are given in Table \ref{tab:model_parameters} and full differential equations are given in Appendix B. 

\subsection{Brief element cycle descriptions}

The model Earth-system N cycle is as follows. Atmospheric N$_2$ dissolves in the ocean, where it can be fixed (i.e., breaking the N triple bond) by bacteria. Fixed N then cycles biologically, and is released as waste or when organisms die. In oxygenated water, this reduced biologic N is quickly nitrified (NH$_4^+$ to NO$_3^-$) by bacteria; in anoxic water it remains as NH$^+_4$. Some organic material sinks into the deep ocean, where most gets remineralized into either NH$^+_4$or NO$^-_3$ depending on O$_2$ levels, and a small portion sinks to the sediments.  In the sediments, organic matter breaks down and bonds as NH$^+_4$ into clays and other K-bearing minerals. Some N also gets incorporated into oceanic crust during hydrothermal alteration. Sediments and oceanic crust get subducted, with a portion of N going into the mantle, a portion outgassed to the atmosphere, and a portion incorporated into arc-generated crust. Mantle N can be outgassed at mid-ocean ridges. Continental crust N, organic or inorganic, can be weathered and added back to the ocean. 

Phosphate enters the ocean due to continental weathering and from mid-ocean ridge outgassing. In the shallow ocean, it is consumed during primary production, both that based on already fixed N and that based on fixing new N. It is then exported to the deep or lower shelf ocean, where it is either remineralized or buried in sediments. Sedimentary and altered oceanic crust PO$^{3-}_4$  can be subducted. A portion goes into the mantle, and a portion into the continental crust. 

The model cycles for K and Ar are only affected by physical, non-biologic processing. All isotopes of both elements mix between ocean boxes, and$^{40}$K decays to$^{40}$Ar in every box where it is found. Continental weathering releases K to the shelf ocean, and Ar to the atmosphere. Air sea gas exchange moves Ar from the surface ocean boxes to the atmosphere. Both species can get incorporated into oceanic crust during hydrothermal alteration, and K can be incorporated into sediments. Again, during subduction, some K and Ar is transported to the mantle, and the remainder either goes to the continental crust (K) or the atmosphere (Ar). Both are added to the deep ocean during mid-ocean ridge outgassing.

\begin{table*}
\begin{center}
\caption[All fluxes contained in model.]{Full list of fluxes used in model. Shown are flux symbol, brief description, and which model species are affected by each flux. In ocean boxes (low, high, shelf, deep), all species includes NO$^-_3$ and NH$^+_4$, while in geologic boxes, all species just includes moles of N.}

\begin{tabular}[h]{ l c c }  
\hline\\
\textbf{Flux} & \textbf{Description} & \textbf{Species affected} \\ 
\hline 
$F_\text{rd}$&  radioactive decay   &$^{40}$Ar, $^{40}$K\\
$F_\text{as}$& air-sea gas exchange & $^{40}$Ar, $^{36}$Ar,  N$_2$ \\
$F_\text{mix}$&water-mixing & all species \\
$F_\text{nit}$&nitrification& NO$^-_3$, NH$^+_4$\\
$F_\text{denit}$&denitrification& NO$^-_3$, N$_2$\\
$F_\text{oldN}$&production from fixed N& NO$^-_3$, NH$^+_4$, PO$^{3-}_4$ \\
$F_\text{newN}$&production from newly fixed N& N$_2$, PO$^{3-}_4$\\
$F_\text{out}$&export from shallow ocean boxes& NO$^-_3$, NH$^+_4$, N$_2$, PO$^{3-}_4$ \\
$F_\text{remin}$&remineralization in shelf and deep&  NO$^-_3$, NH$^+_4$, PO$^{3-}_4$\\
$F_\text{bur}$&burial in sediments& N, PO$^{3-}_4$ \\
$F_\text{seddif}$&diffusion into sediments, out of water& NH$^+_4$\\
$F_\text{sub}$&subduction of crust and sediments&  all species \\
$F_\text{subT}$&total subduction& all species \\
$F_\text{subnet}$&net subduction & all species\\
$F_\text{hydro}$&hydrothermal alteration& all species\\
$F_\text{w}$&weathering&all species \\
$F_\text{cg}$&continental growth& all species\\
$F_\text{scg}$&continental growth from shelf sediments& all species\\
$F_\text{ogarc}$&outgassing-arcs&all species \\
$F_\text{ogmor}$&outgassing-mid ocean ridge&all species \\
\hline\\							
		
\label{tab:fluxes}
\end{tabular}
\end{center}
\end{table*}

	\subsection{$^{40}$K-decay}
	Radioactive decay of$^{40}$K produces$^{40}$Ar in all boxes where$^{40}$K is present: 
	\begin{linenomath*}
	\begin{equation}
	F_\text{rd}^i = \lambda X_\text{40K} R_\text{40K}^i 
		\end{equation}
	\end{linenomath*}
	for $i =$\{all ocean and geologic boxes\}  where $X_\text{40K}$ is the proportion of decays that form$^{40}$Ar and $\lambda$ is the decay constant of$^{40}$K. 
	
	\subsection{Atmosphere}
	The atmosphere contains N$_2$, O$_2$, and both isotopes of Ar.  N$_2$,$^{40}$Ar, and $^{36}$Ar exchange with the ocean following stagnant-lid gas exchange \citep{Liss_and_Slater_1974}, with fluxes positive in the direction of sea to air:  
	
	\begin{linenomath*}
	\begin{equation}
	F_\text{as}^i =  u_{\text{l},j} A^i \left(C_j^i -  H_j P_j \right)
		\end{equation}
	\end{linenomath*}
\noindent	for boxes $i=\{\text{low, high, shelf}\}$ oceans and species $j=\{\text{N}_2,^{40}\text{Ar}, ^{36}\text{Ar}, \text{O}_2\}$.  $u_{\text{l},j} $ is piston velocity in m yr$^{-1}$ calculated by dividing the diffusion constant ($D_i$) by thickness of stagnant boundary ($z_\text{film}$); $A^i$ is ocean box surface area (m$^2$); $C_i^j$ is concentration of species $j$ in box $i$; $H^j$ is Henry's law constant for species $j$ (mol L$^{-1}$ atm$^{-1}$); and $P^j$ is partial pressure of gas $j$, calculated as:   
	\begin{linenomath*}
	\begin{equation}
	P_j = \frac{R^\text{atm}_j}{n_\text{a}}
		\end{equation}
	\end{linenomath*}

\noindent	where $n_\text{a}$ is the number of moles corresponding to one atmosphere pressure (Table \ref{tab:model_parameters}).
	
	Gasses are also added to the atmosphere via outgassing at arcs and both isotopes of Ar are added from continental crust weathering (Sec. \ref{sec:geoflux}). 
	
	Oxygen levels are prescribed. Levels start at $10^{-6}$ present atmospheric level (PAL $=2\times10^{19}$ mol)  through the Hadean and Archean. . Atmospheric O$_2$ increases to 0.01 PAL over 100 Myr at 2.4 Ga, then to 0.8 PAL at 0.630 Ga  (beginning of the Ediacaran). Oxygen rises to modern levels at 0.418 Ga (Devonian). Surface ocean O$_2$ concentrations are calculated assuming a Henry's law equilibrium. 
	
	\subsection{Ocean}
	The ocean is divided into four boxes: high- and low-latitude shallow ocean, shelf ocean, and deep ocean. All species in the model exist in the ocean. There are a number of fluxes, both physical and biogeochemical that occur. Some can occur in all boxes, and some only in specific boxes. Broadly, primary production occurs in shallow ocean boxes,  remineralization and burial occur in deep and shelf boxes, and nitrification and denitrification occur in all boxes. 
	
	\subsubsection{All boxes} 
	
	There is physical water mixing between all ocean boxes. Mixing out of an ocean box is simply the product of concentration of species $j$ in box $i$ and the sum of the water fluxes ($\Delta^{i-i*}$, in L yr$^{-1}$) between box $i$ and all other boxes ($i*$). Mixing in to box $i$ is the sum of the product of the concentrations of species $j$ in each other box ($i*$) and the water flux between box $i$ and each other box ($\Delta^{i-i*}$):	
	
	\begin{linenomath*}
	\begin{equation} 
	F_{\text{mix},j}^i = -C_j^i\sum_{i*}\Delta^{i-i*} +  \sum_{i*}C_j^{i*}\Delta^{i-i*}
		\end{equation}
	\end{linenomath*}

\noindent If, for example, $i=$\{low\}, then $i*=$\{high, shelf, deep\}, and $\Delta^{i-i*}$ is mixing between low and high, low and shelf, and low and deep boxes.

Nitrification occurs in all boxes. The rate of nitrification is dependent on O$_2$ and NH$^+_4$concentrations: 

\begin{linenomath*}
	\begin{equation}
v_\text{Onit}^i = \frac{C_\text{O2}^i}{C_\text{O2}^i+K_\text{Oni}}
	\end{equation}
	\end{linenomath*}

\begin{linenomath*}
	\begin{equation}
v_\text{NOnit}^i = \frac{C_\text{NH4}^i}{C_\text{NH4}^i+K_\text{Nni}}
	\end{equation}
	\end{linenomath*}

The full equation can be parameterized as: 

\begin{linenomath*}
	\begin{equation}
F_\text{nit}^i = \mu_\text{NH4} v_\text{Onit}^i v_\text{NHnit}^i R^i_\text{NH4}
	\end{equation}
	\end{linenomath*}
for box $i=$\{low, high, shelf, deep\}, and where $K_\text{Oni}$ is the half-saturation uptake concentration of O$_2$ used in nitrification ($20~\mu$M), $K_\text{Nni}$ is half-saturation uptake concentration of O$_2$ used in nitrification ($100~\mu$M), and  $\mu_\text{NH4}$ is a rate constant (yr $^{-1}$).
	
	Denitrification can also occur in every ocean box ($i=$\{low, high, shelf, deep\}). It has been observed to follow first-order rate kinetics, with a half-saturation NO$^-_3$ concentration ($K_\text{d}$) of 8 $\mu$M \citep{Goering_1985,Evrard_et_al_2013}:  
	
	\begin{linenomath*}
	\begin{equation}
	v_\text{NOde}^i= \frac{C_\text{NO3}^i}{C_\text{NO3}^i+K_\text{d}}
			\end{equation}
	\end{linenomath*}

	In contrast to nitrification, denitrification only occurs at low O$_2$ levels \citep{Crowe_et_al_2012, Dalsgaard_et_al_2014}, herein parameterized as a Michaelis-Menten type reaction: 
	\begin{linenomath*}
	\begin{equation}
	v_\text{Ode}^i =  \frac{C_\text{O2}^i}{C_\text{O2}^i+O_\text{inhib}}
		\end{equation}
	\end{linenomath*}	
	
	\noindent where $O_\text{inhib}$ is 205 nM \citep{Dalsgaard_et_al_2014}. Thus, we parameterize denitrification as:
	\begin{linenomath*}
	\begin{equation}
      F_\text{den}^i = (1-v_\text{Ode}^i)  v_\text{NOde}^i R^i_\text{NO3}
		\end{equation}
	\end{linenomath*}

		\subsubsection{Shallow ocean}
		Primary production occurs in all shallow ocean boxes ($i=$\{low, high, shelf\}). There are two types of productivity \citep{Fennel_et_al_2005}: that based on  already fixed N ($F_\text{oldN}$) and that based on newly-fixed N$_2$ ($F_\text{newN}$). These two fluxes together  equal total export production ($F_\text{ExT}^i$). The proportion of $F_\text{oldN}$ to $F_\text{newN}$ depends, in part, on the N:P ratio, which we assume will always move towards Redfield (i.e., 16:1). We parameterize this relationship as: 
		
\begin{linenomath*}
	\begin{equation}
 L_\text{R}^i = \frac{N/P^i}{N/P^i+K_\text{NP}}
	\end{equation}
	\end{linenomath*}		

\noindent where $K_\text{NP}$ is 8. 

In addition, $F_\text{oldN}$ can be PO$^{3-}_4$ or N-limited:
\begin{linenomath*}
	\begin{equation}
 L_\text{N}^i = \frac{N_\text{all}^i}{N_\text{all}^i+K_\text{N}}
	\end{equation}
	\end{linenomath*}

\begin{linenomath*}
	\begin{equation}
 L_\text{P}^i = \frac{C_\text{PO4}^i}{C_\text{PO4}^i+K_\text{P}}
	\end{equation}
	\end{linenomath*}

\noindent where $K_\text{N}$ and $K_\text{P}$ are the half saturation uptake values for N and PO$^{3-}_4$. $N_\text{all}^i$ is the sum of $C_\text{NH4}^i$ and $C_\text{NO3}^i$ in box $i$. We also assume productivity ($P$) was 1000 times less efficient before the evolution of oxygenic photosynthesis:
\begin{linenomath*}
	\begin{equation}
 \frac{P}{P_o} =  ((1+1\times10^{-3})+\text{tanh}((t-\text{Phot}_\text{On})/0.1))\times0.5;
	\end{equation}
	\end{linenomath*}
\noindent where $P_o$ is modern primary productivity efficiency (1) and $\text{Phot}_\text{On}$ is the age of oxygenic photosynthesis evolution. 

Thus, $F_\text{oldN}^i$ is: 
\begin{linenomath*}
	\begin{equation}
 F_\text{oldN}^i =  \frac{P}{P_o} L_\text{R}^i  L_\text{N}^i  L_\text{P}^i  R_\text{PO4}^i / \tau_\text{prod}
	\end{equation}
	\end{linenomath*}

\noindent for boxes $i=$\{low, high, shelf\} and where $\tau_\text{prod}$ is a production timescale, set to 0.5 yr. 

Production from N$_2$ fixing is PO$^{3-}_4$-limited, but also depends on N-concentrations and partial pressure of N$_2$. There is a Michaelis-Menten relationship to partial pressure \citep{Klinger_et_al_1989}:

\begin{linenomath*}
	\begin{equation}
v_\text{fix} =\frac{R_\text{N2}^\text{atm}}{R_\text{N2}^\text{atm}+K_\text{f}};
	\end{equation}
	\end{linenomath*}

\noindent where $K_\text{f}$ is $9.87\times10^{18}$ moles, or about the equivalent of 50 mbar pressure. 

Thus, total production based on newly-fixed N$_2$ is: 

\begin{linenomath*}
	\begin{equation}
\begin{split}
 F_\text{newN}^i =   \frac{P}{P_o} (1-L_\text{R}^i) L_\text{P}^i   v_\text{fix} R^i_\text{PO} /\tau_\text{prod}
 \end{split}
	\end{equation}
	\end{linenomath*}

We note that we do not include abiotic N-fixing in our model \citep[e.g.,][]{Navarro_et_al_2001}. Total export production ($F_{\text{ExT}}^i$) from boxes $i=$\{low, high, shelf\} is the sum of $F_\text{oldN}^i$ and $F_\text{newN}^i$. Export production  in both the low- and high-latitude boxes  go to the deep ocean, and shelf export stays in the shelf box. Phosphate loss ($F_{\text{out,PO4}}^i$) from shallow boxes is equal to $F_{\text{ExT}}^i$. Fixed NO$^-_3$ and NH$^+_4$losses are equal to:
\begin{linenomath*}
	\begin{equation}
 F_{\text{out},j}^i = 16F_\text{oldN}^i\frac{ C_j^i}{N_\text{all}^i} 
	\end{equation}
	\end{linenomath*}
 \noindent for $j=$\{NO$^-_3$, NH$^+_4$\} and N$_2$ loss is: 
 \begin{linenomath*}
	\begin{equation}
 F_\text{out,N2}^i = 16 F_\text{newN}^i
	\end{equation}
	\end{linenomath*}

\noindent as we keep track of moles N, not moles N$_2$. 

\subsubsection{Shelf ocean, deep ocean, reactive shelf and deep sediments}
	PO$^{3-}_4$ and N can be remineralized or buried, and K can be buried in sediments. The efficiency of remineralization depends on oxygen concentration:
	
	\begin{linenomath*}
	\begin{equation}
 f_\text{aerobic}^i = \frac{C_\text{O2}^i}{C_\text{O2}^i + K_\text{O}}
  	\end{equation}
	\end{linenomath*}
  
  \begin{linenomath*}
	\begin{equation}
 F^i_\text{reminfrac, j} = 0.9 f_\text{aerobic}^i + 0.7(1-f_\text{aerobic}^i). 
 	\end{equation}
	\end{linenomath*}
 
\noindent for boxes $i=\{\text{shelf ocean, deep ocean, reactive shelf sediments, reactive deep sediments} \}$ and species $j=\{\text{PO}^{3-}_4, \text{NH}^+_4\}$ and where $K_\text{O}$ is 50 $\mu$M. The values in the above equations are tuned to have aerobic remineralization convert $90\%$ of export production, while only $70\%$ gets remineralized under anaerobic conditions. Nitrogen remineralization and burial is assumed to be in Redfield ratio (i.e.,16:1) compared to PO$^{3-}_4$.  Nitrogen is remineralized as NH$^+_4$.  
 
  	Export production from the high and low-latitude shallow boxes is remineralized in the deep ocean and export production in the shelf ocean is remineralized in the shelf ocean. That which is not remineralized in the deep or shelf ocean boxes goes to the reactive deep and shelf sediments, respectively. Remineralization can occur in these reactive sediment boxes, with anything not remineralized buried to the non-reactive deep and shelf sediments.   Additionally, nitrification, denitrification, and diffusion occur in the reactive sediment layers. Diffusion ($F_\text{diff}^i$) is parameterized as: 

\begin{linenomath*}
	\begin{equation}
F_{\text{diff},j}^i = D_j \left(C_{\text{O2},j}^{i*} - \frac{C_{\text{O2},j}^i}{D_\text{length}} \right) SA^{i*}
	\end{equation}
	\end{linenomath*}

for boxes $i=\{\text{Reactive shelf sediments, reactive deep sediments} \}$ and $i*=\{\text{shelf ocean,  deep ocean} \}$. $D_j$ is the diffusion constant and $D_\text{length}$ is diffusive length, which is set to 0.01 m.

\captionsetup{width=\textwidth}
\begin{center}
\begin{longtable}{l c r c}
\caption{All model constants, shown with references.  References are: H14 \citep{Haynes_et_al_2014}, S99 \citep{Sander_1999}, F05 \citep{Fennel_et_al_2005}, LS74 \citep{Liss_and_Slater_1974}, B16 \citep{Bristow_et_al_2016}, D14 \citep{Dalsgaard_et_al_2014}, G85 \citep{Goering_1985},   E13 \citep{Evrard_et_al_2013},  K89 \citep{Klinger_et_al_1989},      JG15 \citep{Johnson_and_Goldblatt_2015}, W10 \citep{Winter_2001}, and K10 \citep{Korenaga_2010}}\\

\hline \textbf{Parameter} & \textbf{Definition} & \textbf{Value} & \textbf{Reference} \\ \hline
\endfirsthead

\multicolumn{4}{c}%
{{\bfseries \tablename\ \thetable{} Model Constants -- continued from previous page}} \\ 

\hline \textbf{Parameter} & \textbf{Definition} & \textbf{Value} & \textbf{Reference} \\ \hline

\endhead

\hline \multicolumn{4}{r}{\textit{Continued on next page}} \\ \hline
\endfoot

\endlastfoot
	$\lambda$& decay constant of$^{40}$K (yr $^{-1}$) & $5.543\times10^{-10}$ yr $^{-1}$ & H14\\
	$X_\text{40}$& fraction of decay producing $^{40}$Ar & $0.1072$ & H14 \\	
	$D_\text{N$_2$}$ & Diffusion constants  (cm$^2$ s$^{-1}$ ) & $1.88\times10^{-5}$  & H14\\ 
	$D_\text{Ar}$  &   & $1.88\times10^{-5}$  & H14 \\ 
	$D_\text{O$_2$}$ & & $2.5\times10^{-5}$ & H14\\ 
	$D_\text{length}$ & Diffusion length & 0.01 m & \\
	$H_\text{N$_2$}$ & Henry's law constants (mol L$^{-1}$ atm $^{-1}$) &  $1.3\times10^{-3}$ &S99 \\
	$H_\text{Ar}$ &    & $6.1\times10^{-4}$&S99  \\ 
	$H_\text{O$_2$}$ & & $6.1\times10^{-4}$  & S99 \\ 
	$n_a$& moles equal to one atm pressure & $1.72\times10^{20}$ & this study\\
	$d$& ocean depth (m) & 4500  & F05\\
	$z_{film}$& stagnant lid thickness (m) & $1.5\times10^{-5}$  & LS74 \\ 
	$A^\text{high}$& ocean box surface area (m$^2$) & $2.3\times10^{12}$ & F05\\
	$A^\text{low}$& & $2.3\times10^{12}$& F05\\
	$A^\text{shelf}$& & $5.1\times10^{11}$ & F05\\
	$V^\text{high}$& ocean box volumes (L) & $9\times10^{18}$ & F05\\
		$V^\text{low}$& & $9\times10^{18}$&F05\\
	$V^\text{shelf}$& & $1.5\times10^{18}$&F05\\
	$V^\text{deep}$& & $8.1\times10^{20}$ &F05\\
	$\Delta^\text{low-high}$ & mixing between ocean boxes (Sv) &30 & F05\\
	$\Delta^\text{low-shelf}$ &  & 30  &F05\\
	$\Delta^\text{low-deep}$ & & 30 & F05\\
	$\Delta^\text{high-shelf}$ & & 30 & F05\\
	$\Delta^\text{high-deep}$ & & 50 & F05 \\
	$\Delta^\text{shelf-deep}$ & &5-100 &This study\\
	$K_\text{Oni}$&half-saturation uptake of O$_2$ in nitrification & 283 nM& B16 \\
	$K_\text{Nni}$&half-saturation uptake of NH$^+_4$in nitrification & 100 $\mu$M& F05 \\	
	$K_\text{d}$& half-saturation uptake of NO$^-_3$ in denitrification& $8~\mu$M &  \\
	$K_\text{NP}$& Redfield term & $8.47~\mu$M &G85, E13 \\
	$K_\text{N}$& half-saturation uptake of total N&$1.6~\mu$M & F05\\
	$K_\text{P}$& half-saturation uptake of PO$^{3-}_4$ &$0.1~\mu$M & F05\\
	$K_\text{f}$& half-saturation uptake of N$_2$ during fixing& $9.87\times10^{18}$ mol& K89\\
	$K_\text{O}$&  O$_2$ concentration term & $50 ~\mu$M & \\
	$K_\text{dist}$& percent of species extracted & 0.90 & \\ 
	 &during MORB-genesis & & \\
	$\tau_\text{prod}$ & Export production time-scale & 0.5 yr$^{-1}$ & \\
	$\tau_\text{W}$&rate constant of weathering  & $3.33\times10^{-9}$ yr $^{-1}$  & \\
	$\tau_\text{scg}$& rate constant of & $1\times10^{-8}$  yr $^{-1}$ & this study\\
	$\mu_\text{NH4i}$& nitrification rate constant & 1 yr$^{-1}$ & F05 \\	
	O$_\text{inhib}$ &Oxygen inhibition for denitrification & 205 nM & D14 \\
	$u_{co}$ & Modern spreading rate & 0.05  m yr$^{-1}$ & K10\\
	$T_o$ & Modern mantle potential temperature & 1350 $^{\circ}$C &  K10 \\
	$\rho_c$ & Ocean crust density & 3000   kg m$^{-3}$ \\ [0.5ex]		
	$M^\text{desed}$& mass of deep sediments  & $7.3\times10^{23}$ g & JG15\\
	$M^\text{occrust}$&mass of ocean crust & $5.4\times10^{24}$ g &JG15\\
	$M^\text{mantle}$&mass of mantle  & $4\times10^{27}$ g& JG15\\
		& shelf sediments to cont. crust   & & \\
	$a_\text{W}$& Anoxic weathering fraction & $0.1$   & this study\\
	$k_\text{weath}$& Weathering rate O$_2$ dependence & $1\times10^{-3}$   & this study\\
	V$_\text{hydro}$&  hydrothermal circulation volume & $1.6$ Sv   & this study\\
	H$\text{eff}$ & Hydrothermal retention efficiency & 0.1-1, varies by species & this study \\
	$P_m$& percent partial melt  & $10\%$ & W01 \\
		& during MORB-genesis & & \\
	$S$& spreading rate (m yr$^{-1}$) & varies & K10 \\ 
	$h_\text{sed}$& sediment thickness  & 500 m& this study \\ 
	$h_c$ & Ocean crust thickness & 16000 to 8000  m &  K10\\
	$L_s$& length of subduction zones  &$4\times10^7$ m& this study \\
	$L_r$& length of mid-ocean ridges  &$8\times10^7$ m& this study \\
	$f_\text{shsed}$ & fraction shelf sediments subducted & $1\times10^{-9}$  & this study \\
	$f_\text{Ncgarc}$ & fraction N added to continental crust & 0.5  & this study \\	\hline	
\label{tab:model_parameters}

\end{longtable}
\end{center}

	\subsection{Geologic model}
	\label{sec:geoflux}
	The model is driven by a mantle cooling history from \citet{Korenaga_2010} and \citet{Padhi_et_al_2012}. This model suggests that mantle temperatures ($T_m$) increased through the early Archean, reached their peak in the middle Archean, and have been decreasing to the modern day (Table \ref{tab:mantle_evo}, Fig. \ref{fig:model_forcings}). Heat flux ($Q$) followed a distinct evolution, reaching its maximum later than the mantle temperature apex (Table \ref{tab:mantle_evo}). Temperature and heat flux are used to parameterize a plate velocity ($u_c$):
	
	\begin{linenomath*}
	\begin{equation}
	u_c = u_{co} \frac{Q}{Qo} \left(\frac{T_o}{T_m}\right)^2
		\end{equation}
	\end{linenomath*}
	where $u_{co}$, $Q_o$, and $T_o$ are modern plate velocity ($0.05$ m yr$^{-1}$), heat flux (39 TW), and average mantle temperature (1350 $^{\circ}$C). 
	
	The model then  calculates  crust production at mid-ocean ridges by combining spreading rate with ridge length ($L_r$, m) and crust thickness ($h_c$, m): 
	
	\begin{linenomath*}
	\begin{equation}
	O_p  = u_c h_c L_r  \rho_c
		\end{equation}
	\end{linenomath*}
	
\noindent	where $\rho_c $ is crust density (kg m$^{-3}$), and $h_c$ decreases linearly through time \citep{Sleep_and_Windley_1982} from 16 km at the beginning of the model to 8 km at $t=4.5$ Gyr. We assume that the amount of crust subducted ($O_s$) is equal to $O_p$. 

 \begin{center}
\begin{table*}
\caption[Mantle temperature and heat flux evolution.]{Mantle temperature ($T_m$) history, heat flux ($Q$), and spreading rate ($u_c$) evolution from \citet{Korenaga_2010, Padhi_et_al_2012}.}
\centering
\begin{tabular}[h]{ l c c c}  
\hline\\

	\bf{Age (Ga)}	&	\bf{$T_m~^{\circ}$C } &\bf{$Q$ (TW)} & \textbf{$u_c$ (cm yr$^{-1}$)}	 \\[2ex] \hline
	0	 & 1350 & 39 & 5.00\\
	0.5	 & 1425 & 43&5.55\\
	1.0	 & 1490 & 41&4.68 \\
	1.5	 & 1540 & 40& 4.22 \\
	2.0	 & 1600 & 39& 3.75 \\
	2.5	 & 1680 & 38& 3.28 \\
	3.0	 & 1700 & 37& 3.05 \\
	3.5	 &1700 & 37& 3.05 \\
	4.0	 &1670 & 37.5& 3.23 \\
	4.5	 &1650 & 38 & 3.38 \\
	 [0.5ex]
\hline\\

\label{tab:mantle_evo}
\end{tabular}
\end{table*}
\end{center}

	\subsection{Sediments}
		
		Anything that does not get remineralized in the reactive sediment boxes gets buried ($F_\text{bur}^i$) in sediments: 
		\begin{linenomath*}
	\begin{equation}
		F_\text{bur}^i = F_\text{burial}^{i*} - F_\text{remin}^{i*}
		\label{eq:burial}
			\end{equation}
	\end{linenomath*}
		
		\noindent for boxes $i=\{\text{shelf sediments, deep sediments}\}$ and $i*=\{\text{reactive shelf sediments, reactive deep sediments}\}$. 
		
Species are subducted from both deep and shelf sediments. A constant fraction of shelf sediment species gets subducted,:
		\begin{linenomath*}
	\begin{equation}
		 F_\text{sub,j}^\text{shsed} = f_\text{shsed}R_j^\text{shsed}
		 	\end{equation}
	\end{linenomath*}
		
		\noindent where $f_\text{shsed}$ is the fraction of shelf sediments that subduct ($1\times10^{-9}~\text{yr}^{-1}$).  Deep sediments subducted are equal to: 
		
		\begin{linenomath*}
	\begin{equation}
		F_\text{sub,j}^\text{dsed} = \frac{m_\text{sub}^\text{dsed} R_j^\text{dsed}}{M^\text{dsed}}
			\end{equation}
	\end{linenomath*}
		where $m^\text{desed}$ is the mass of deep sediments subducted and $M^\text{dsed}$ is the mass of deep sediments ($7.4\times10^{23}$ g). Mass of sediments subducted is: 
		\begin{linenomath*}
	\begin{equation}
		m_\text{sub}^\text{dsed} = L_\text{s} S h_\text{sed} \rho_\text{sed}
			\end{equation}
	\end{linenomath*}
		where $L_s$ is subduction zone length (m), $S$ is spreading rate (m yr$^{-1}$) calculated from Korenaga model, $h_\text{sed}$ is thickness of sediments ($500$ m), and $\rho_\text{sed}$ is sediment density (2.5 g cm$^{-3}$). 
		
		In addition, shelf sediments have a residence time of $100$ Myr, or a rate constant of $\tau_\text{scg} = 1\times10^{-8}~\text{yr}^{-1}$. Shelf sediments are added to the continental crust, representing a proxy for continental growth by collision and accretion:
		
		\begin{linenomath*}
	\begin{equation}
		F_\text{scg,j} = \tau_\text{scg} R_j^\text{shsed}
			\end{equation}
	\end{linenomath*}
		
		\noindent for $j=$\{all species\}.  
	\subsection{Crust}
	
	\subsubsection{Oceanic}
	
Species can enter the oceanic crust through hydrothermal alteration, and the leave the oceanic crust during subduction. We envision hydrothermal processes essentially as serpentinization, and overall it adds N to the oceanic lithosphere \citep{Halama_et_al_2014}. The amount of hydrothermal alteration is related to both speciation and a volume of hydrothermal fluid flow per year:

		\begin{linenomath*}
	\begin{equation}
		F_\text{hydro, j} = C_j^\text{deep} V_\text{hydro} H_\text{eff}
			\end{equation}
	\end{linenomath*}
	\noindent for all model species. $H_\text{eff}$ is 1 for K and NH$^+_4$, 0.5 for NO$^-_3$ and PO$^{3-}_4$, and 0.01 for Ar. We set $V_\text{hydro}$ equal to 1.6 Sv  \citep{Elderfield_and_Schultz_1996, German_and_Seyfried_2014} for nominal runs, but allowed it to vary during sensitivity tests.  
			
The subduction flux is calculated by multiplying the mass of crust subducted per year  by each species concentration in the crust:
		\begin{linenomath*}
	\begin{equation}
F_{\text{sub},j}^\text{oc} = \frac{R_j^\text{ocrust}O_p}{M^\text{occrust}}
	\end{equation}
	\end{linenomath*}

\noindent where $O_p$ is ocean crust produced and $M_\text{occrust}$ is the total mass of crust (g).
		
		Thus, the total amount of each species subducted ($F_\text{subT}$) is 
\begin{linenomath*}
	\begin{equation}
F_{\text{subT},j} = F_{\text{sub},j}^\text{oc} + F_{\text{sub},j}^\text{dsed} + F_{\text{sub},j}^\text{shsed}
	\end{equation}
	\end{linenomath*}

Subducted species will either be driven off the slab and sediments or carried beyond the subduction barrier and into the mantle.  The proportion that is driven off the slab  is determined by mantle temperature: higher temperature means less material goes into the mantle, and lower temperature means more material goes into the mantle. Subducted fraction is calculated from an average geothermal gradient ($G_\text{sub}$ in $^\circ$C km$^{-1}$), which in turn is calculated from an average mantle temperature ($T_m$): 

\begin{linenomath*}
	\begin{equation}
G_\text{sub} = \frac{12.2(T_m-273)}{2900}
	\end{equation}
	\end{linenomath*}

\noindent where $T_m$ is in kelvin, $2900$ is mantle depth in km, and 12.2 is a conversion factor to adjust average mantle temperature, consistent with the modern average mantle geothermal gradient. Subducted fraction is a hyperbolic tangent fit to the data from modern geothermal gradients and subducted fluxes at three modern subduction zones  \citep{Elkins_et_al_2006, Mitchell_et_al_2010, Zelenski_et_al_2012}, and can vary between 0.1 and 1:
\begin{linenomath*}
	\begin{equation}
f_\text{sub} = 0.5 \left(1.1-0.9 \tanh\left(\frac{G_\text{sub} -6}{0.6}\right)\right)
	\end{equation}
	\end{linenomath*}
The values inside the tanh parenthetical, 6 and 0.6, have units of $^\circ$C km$^{-1}$.  We again note previous work that has indicated there is likely more complication in the ratio between N that is subducted at the trench and that which is sequestered to the mantle. While temperature is assumed to have a first-order effect in our model, redox \citep{Libourel_et_al_2003,Li_et_al_2016}, pH \citep{Mikhail_and_Sverjensky_2014}, and distribution between fluids and melt \citep{Li_et_al_2015, Mallik_et_al_2018}, may all have effects which are not considered here.

Thus, the flux of species subducted to the mantle is the product of subducted fraction, concentration in sediments or crust, and mass of sediments/crust subducted per year:

\begin{linenomath*}
	\begin{equation}
F_\text{subnet,j} = f_\text{sub} F_\text{subT,j}.
	\end{equation}
	\end{linenomath*}

\noindent for $j=$\{all species\}.

		\subsubsection{Continental}
		
		That which is not subducted will either be outgassed at arcs ($F_\text{ogarc}$) or be incorporated into the continental crust ($F_\text{cg}$). All Ar is outgassed, all K and PO$^{3-}_4$ goes into the continental crust. For N,   $f_\text{Ncgarc}$ is set to 0.5. That is, half of N released from subducted materials is outgassed at arcs and half is incorporated into the continental crust. This value, 0.5, is an assumption in our model. There is very little data concerning N in subduction zones that is released from the slab. Nitrogen isotopes in granites indicate a biologic source \citep{Boyd_2001, Johnson_and_Goldblatt_2017}, which could be from subducted material. More analysis of granitic rocks would help characterize this flux. Thus:
		
		\begin{linenomath*}
	\begin{equation}
		F_{\text{cg},j} = (1-f_\text{sub}) F_\text{subT,j}
			\end{equation}
	\end{linenomath*}
		
\noindent for  $j=$\{K,$^{40}$K,$^{40}$Ar,$^{36}$Ar\} and

	\begin{linenomath*}
	\begin{equation}
		F_{\text{cg,N}} = f_\text{Ncgarc}(1-f_\text{sub}) F_\text{subT,N}
			\end{equation}
	\end{linenomath*}

\noindent for N. All subducted PO$^{3-}_4$ is added to the continental crust. 
		
Species in the continental crust have  a residence time of 300 Myr, or time constant ($\tau_w$) of $3.33\times10^{-9}$ yr$^{-1}$, which is equivalent to half a Wilson cycle \citep{Nance_and_Murphy_2013}. Weathering efficiency depends on atmospheric O$_2$, with weathering increasing with increasing O$_2$:

\begin{linenomath*}
	\begin{equation}
		W_{\text{eff},j} = \tau_{\text{w}} \left( a_\text{W} + (1-a_\text{W}) \frac{PAL_\text{O2}}{PAL_\text{O2} + k_\text{weath}} \right)
			\end{equation}
	\end{linenomath*}

\noindent Where $a_\text{W}$ is the fraction of available material weathered under anoxic conditions (0.1), $PAL_\text{O2}$ is atmospheric O$_2$ compared to present atmospheric levels, and $k_\text{weath}$ is a weathering rate constant ($1\times10^{-3}$).  Weathered$^{36}$Ar and$^{40}$Ar are released to the atmosphere, while all other species (N, PO$^{3-}_4$, K) are added to the shelf ocean. As there is no crustal organic material in the model, all continental N is weathered as NH$^+_4$, which is the mineralogically most stable form of N. Weathering ($F_\text{w}$) is parameterized as:

		\begin{linenomath*}
	\begin{equation}
		F_{\text{w},j} = W_{\text{eff}} R_j^\text{contcrust}
			\end{equation}
	\end{linenomath*}

	\subsection{Mantle}

Species are added to the mantle at subduction zones ($F_\text{subnet}$). It is assumed that they instantly homogenize into the mantle (i.e., there are no separate mantle domains). Species leave the mantle through degassing at mid-ocean ridges. Degassing is  the product of the concentration ($C_j^\text{man}$) of the species in the mantle and the mass of mantle involved in crust genesis ($M_\textrm{melt}$). $M_\textrm{melt}$ is set to $10$ times the mass of oceanic crust produced ($O_{p}$), which and represents generation of crust by $10\%$ partial melt ($P_m$).  We assume $90\%$ of all species are partitioned to the melt during partial melting ($K_\text{dist}$), with $10\%$ remaining in the residual. We have chosen this partition of melt to residual to account for the observation that mantle rocks that have undergone some melting still have low, but measurable N of less than 1 ppm \citep[][and references therein]{Johnson_and_Goldblatt_2015}. Thus, mid-ocean ridge outgassing is:   
	
	\begin{linenomath*}
	\begin{equation}
	F_{\text{ogmor},j} = C_j^\text{man}  P_m O_{p}  K_\text{dist}. 
		\end{equation}
	\end{linenomath*}

We note that there is no explicit treatment of intra-plate, or hot spot, volcanism in the model. In addition, we do not distinguish between the upper mantle, transition zone, and lower mantle. There are redox changes with depth in the mantle \citep[e.g.][]{Frost_and_McCammon_2008}, which have important effects on N solubility in mantle minerals \citep{Li_et_al_2013, Li_et_al_2016}. As discussed in \citet{Li_et_al_2013} and \citet{Johnson_and_Goldblatt_2015}, the mantle likely has an enormous capacity for N, which likely exceeds its actual content at any given time. Future modeling work including mantle structure and redox evolution would be an important addition to the work shown herein. 

\subsection{Details on code structure}

The model code was constructed to prioritize flexibility. Due to the high number of unknowns in the system, giving flexibility was important. We set up the reservoir bookkeeping as a structure array in Matlab. This allows for dynamic field names to be used, which assists in ease of code reading. We also constructed it so that initial conditions are read in through a separate text file. This allows easy changes, but it is also flexible as not every species has to be in every box. It also calculates$^{40}$K from K initial conditions, reducing input time and error. 

The differential equation file is arranged so that it is straightforward to turn various fluxes off and on. The purpose for this design is that this model, or one like it, could be used for not only Earth history, but could be applied to planetary evolution in general. Different planetary evolution pathways may or may not involve subduction, different atmospheric compositions, or differing biologic pathways and metabolisms. Testing the response of the system to such differences,  perturbations, and the presence or absence of one or more fluxes could be  of great value in studying planetary evolution. 

In detail, we used Matlab's ode15s solver. This is a variable-step, variable-order solver that uses numerical differentiation formulas of orders 1 to 5. We set the relative error tolerance to $1\times10^{-7}$ and a maximum step size of $10^6$ years. Code is available in the supplementary material. Please contact us if you wish to use this code in order to obtain the latest version. 

\section{Results and Discussion}
We ran the model in all runs for 4.5 Ga, after a spin up period of 10 Myr to equilibrate atmosphere,  ocean, and sediment boxes. All biologic N fluxes are ``available'' at each model step.

\subsection{Nominal Run}
In order to test the effects of different conditions over Earth history, such as oxygenic photosynthesis evolution time and style of mantle cooling, we first describe the results of a nominal model run.

This realization is based on a conservative set of assumptions regarding initial and boundary conditions and choice of parameterizations (Fig. \ref{fig:model_forcings}).  Mantle cooling and mid-ocean ridge crust production (i.e., mid-ocean ridge outgassing)  is from  \citet{Korenaga_2010}, with the fraction of subducted N retained to the deep mantle dependent on mantle temperature. The atmospheric O$_2$ history is prescribed, and oxygenic photosynthesis evolves at 2.8 Ga. Plate tectonics starts at 3.5 Ga, continental weathering timescale is 300 Myr, and hydrothermal alteration is parameterized as a fixed volume flow (1.6 Sv). 

\begin{figure}
\begin{center}
\includegraphics[width=\textwidth]{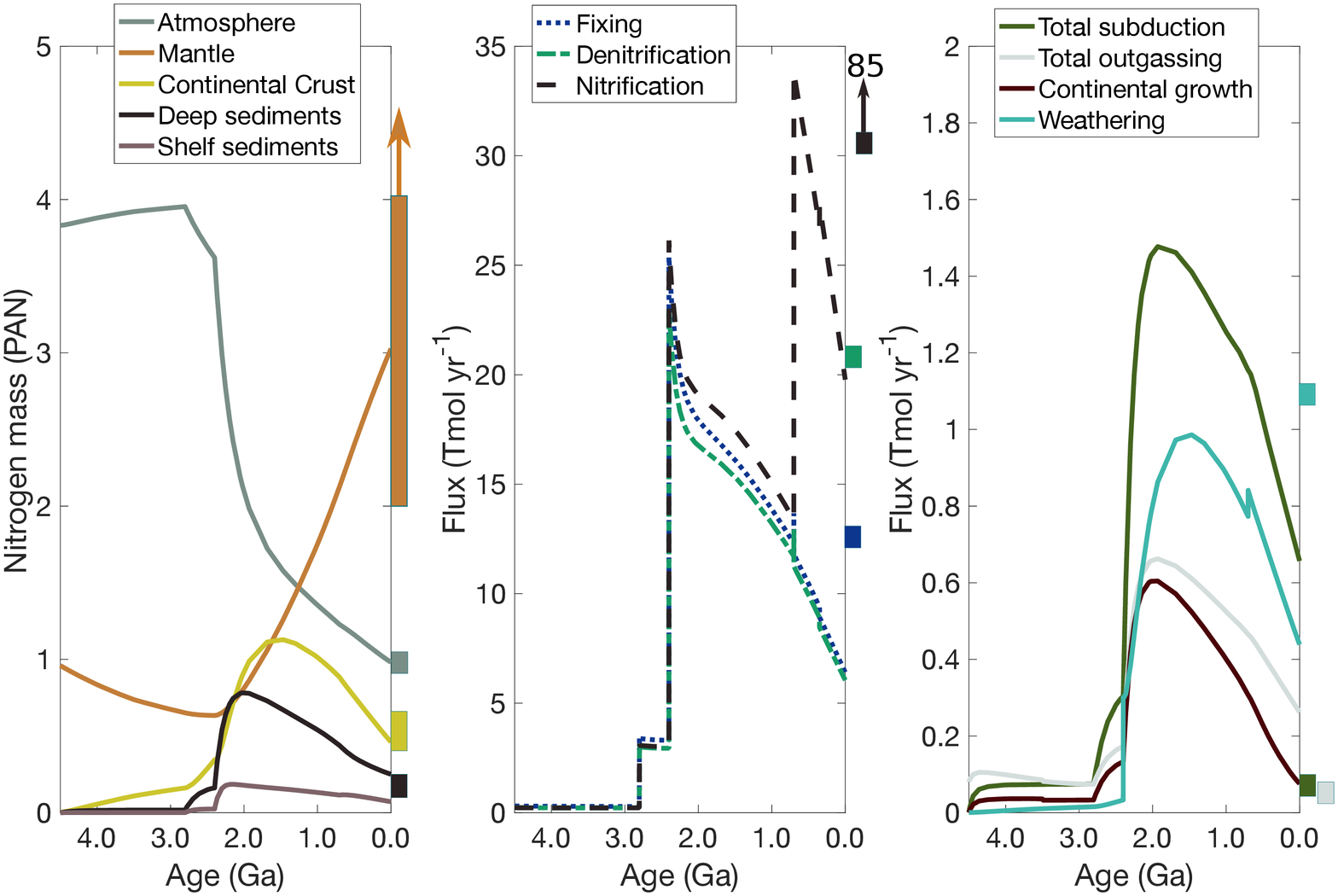}
\caption{a. Nitrogen reservoirs through time from a nominal run. b. Biologic nitrogen fluxes and c. Geologic fluxes. Estimates for modern amounts are shown as colored bars next to each panel \citep{Gruber_2008,  Gruber_and_Galloway_2008, Busigny_et_al_2011, Vitousek_et_al_2013, Halama_et_al_2014, Johnson_and_Goldblatt_2015,  Catling_and_Kasting_2017, Mallik_et_al_2018, Houlton_et_al_2018}.}
\label{fig:standardNresv}
\end{center}
\end{figure}

\begin{figure}[h]
\begin{center}
\includegraphics[width=\textwidth]{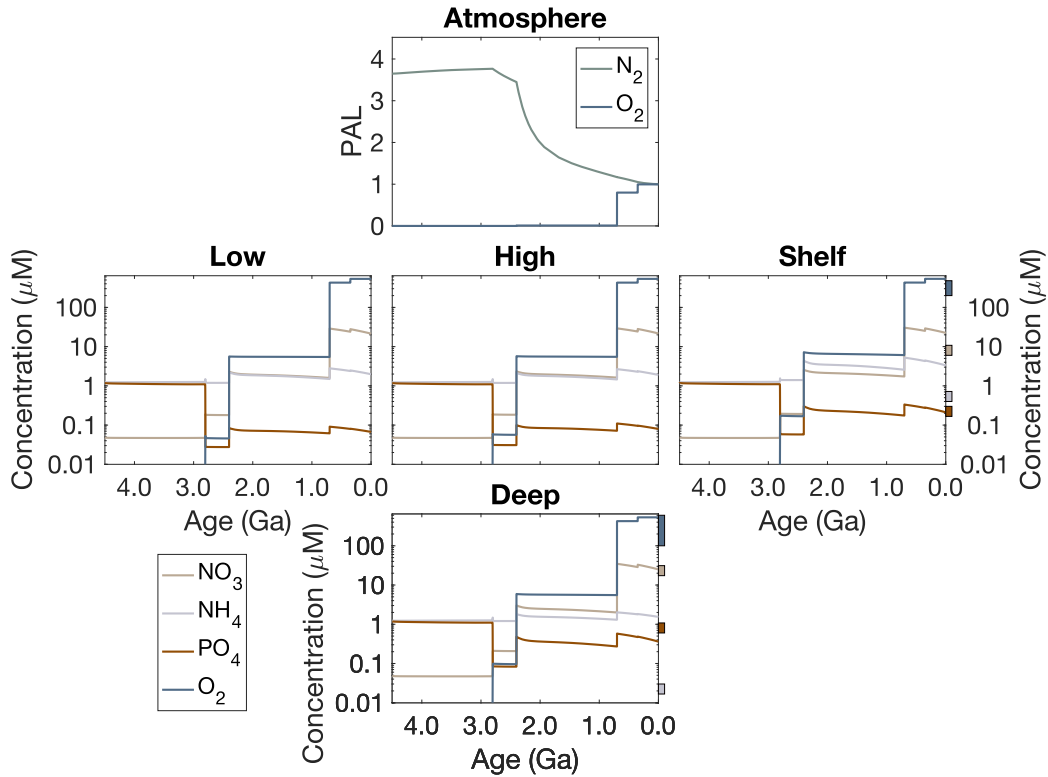}
\caption{Nitrogen, O$_2$,  and PO$^{3-}_4$ surface reservoirs through time for a nominal run. NH$^+_4$is the dominant N species until the Neoproterozoic Oxygen Event, and PO$^{3-}_4$ is high until the evolution of oxygenic photosynthesis. A small O$_2$ oasis exists in the late Archean shelf ocean. Literature estimates for modern values are shown as colored bars next to shallow boxes; the shallow ocean bars are for the average shallow ocean, and deep ocean represents 1000 m depth \citep{Gruber_2008, Garcia_et_al_2014}.}
\label{fig:standardAtmOc}
\end{center}
\end{figure}

 We estimate the proportions of N that start in the atmosphere and the mantle at the end of the magma ocean phase of Earth history, and use this as the initial condition for the nominal run (Appendix \ref{apdx:Initial_N}). Using results from \citet{Libourel_et_al_2003}, which relates pN$_2$ to N dissolved in basaltic magma, and a mantle $f$O$_2$ of IW-2 \citep{Wood_et_al_2006}, which is expected at the end of core formation, we calculate N$_2$ concentration in a magma ocean for a range of atmospheric pN$_2$ values. We assume the entire mantle experienced a magma ocean phase. Then, given this relationship, we can estimate a total N budget and what proportion of that N starts in the atmosphere and the mantle. We select a total N budget for the nominal run to  be consistent with budget estimates from \citet{Johnson_and_Goldblatt_2015}, and one that reproduces the current distribution of N in the atmosphere (1 PAN) and the mantle (>3 PAN). Our starting conditions are therefore total N of 4.8 PAN, with 80\% starting in the atmosphere and 20\% in the mantle. The evolution of major N reservoirs (atmosphere, mantle, continental crust, ocean sediments) is shown in Fig. \ref{fig:standardNresv} with atmosphere-ocean gases and nutrients shown in Fig. \ref{fig:standardAtmOc}.

\captionsetup{width=\textwidth}
\begin{center}
\begin{longtable}{l c c l}
\caption{Nitrogen reservoir and flux comparisons: nominal model output results compared to literature N mass estimates from \citet{Johnson_and_Goldblatt_2015}, with continental crust from \citet{Johnson_and_Goldblatt_2017}. Reservoirs are in units of $10^{18}$ kg N, while ocean concentrations are shown in $\mu$M. Shallow ocean model results are the average of low-latitude, high-latitude, and shelf boxes. Deep ocean values from the literature are for 1000 m depths. Model values are from the end of nominal model run (i.e., modern values). Fluxes are in Tmol N yr$^{-1}$ unless otherwise noted. Model results are shown for modern day reservoir fluxes. }\\ 

\hline \textbf{Reservoir/Flux} & \textbf{Model} & \textbf{Literature}  &\textbf{Reference} \\ \hline
\endfirsthead

\multicolumn{4}{c}%
{{\bfseries \tablename\ \thetable{} Reservoir comparison -- continued from previous page}} \\ 

\hline \textbf{Reservoir/Flux} & \textbf{Model} & \textbf{Literature}  &\textbf{Reference} \\ \hline

\endhead

\hline \multicolumn{4}{r}{\textit{Continued on next page}} \\ \hline
\endfoot

\endlastfoot
Atmosphere & 3.92 & 4 & \citet{Johnson_and_Goldblatt_2015} \\
Mantle & 12 & $24\pm16$&  \citet{Johnson_and_Goldblatt_2015}  \\
Continental Crust & 1.8 & $1.7-2.7$&  \citet{Johnson_and_Goldblatt_2015,Johnson_and_Goldblatt_2017}  \\
Oceanic Lithosphere & 0.05 & $0.2\pm0.02$&  \citet{Johnson_and_Goldblatt_2015}  \\
Total ocean Sediments & 1.2 & $0.41\pm0.2$ &  \citet{Johnson_and_Goldblatt_2015}  \\
Shallow ocean NO$^-_3$ &  22 &  $7$ & \citet{Gruber_2008}  \\
Deep ocean NO$^-_3$ & 25 & $31$ & \citet{Gruber_2008} \\
Shallow ocean NH$^+_4$& 2.4 & $0.3$ & \citet{Gruber_2008} \\
Deep ocean NH$^+_4$& 1.5 &  $0.01$  & \citet{Gruber_2008} \\
Shallow ocean PO$^{3-}_4$ & 0.12 & $<1$ & \citet{Garcia_et_al_2014}\\
Deep ocean PO$^{3-}_4$ & 0.36 &$1-3$ & \citet{Garcia_et_al_2014} \\
Shallow ocean O$_2$ & 533 & $220 - 400$ & \citet{Garcia_et_al_2014} \\
Deep ocean O$_2$ & 530 &$40 - 400$ & \citet{Garcia_et_al_2014} \\ \hline 
\multicolumn{4}{l}{\textit{Biologic fluxes}} \\ \hline
N-fixing  & 6.4  & 13   & \citet{Gruber_and_Galloway_2008,Vitousek_et_al_2013} \\
Nitrification  & 19 & 85 & \citet{Gruber_2008} \\
Denitrification  & 6 & 22 & \citet{Gruber_2008}\\
N-remineralization  & 21 & 93 & \citet{Gruber_2008} \\ [1ex]
\multicolumn{4}{l}{\textit{Geologic Fluxes}}  \\ \hline
Continental weathering    &0.43 & 1.1 & \citet{Houlton_et_al_2018}\\
Burial/sedimentation & 0.73 & 0.07 &  \citet{Gruber_2008} \\
Total subduction  & 0.66 & 0.0064 & \citet{Mallik_et_al_2018} \\
	& &0.0094  & \citet{Busigny_et_al_2011} \\
	& & 0.1 & \citet{Halama_et_al_2014} \\
Arc outgassing  & 0.08 & 0.0375 & \citet{Catling_and_Kasting_2017}\\
Mid-ocean ridge outgassing  & 0.2 & 0.0038 & \citet{Catling_and_Kasting_2017} \\
Total outgassing  & 0.28 & 0.09 & \citet{Catling_and_Kasting_2017} \\
\hline
\label{tab:mod_comparisons}
\end{longtable}
\end{center}

 We focus first on model output at the modern day. The nominal run reproduces the modern atmospheric and estimated mantle N masses well. The mantle value, specifically, is somewhat lower, but  within the estimated mantle N budget from \citet{Johnson_and_Goldblatt_2015}, which is $7\pm4$ PAN. The nominal run has 3.25 PAN in the mantle at modern. We present full comparisons with values from the literature in Table \ref{tab:mod_comparisons}.

Model output is consistent with estimates for the modern day N budget in deep sediments of $\sim 0.1$ PAN \citep{Johnson_and_Goldblatt_2015}, and  continental crust. For example, recent work using glacial tills as a proxy for upper continental crust through time suggest a secular increase in crustal N  during the Precambrian \citep{Johnson_and_Goldblatt_2017}. The authors suggest that isotopic evidence is most consistent with this addition of N to the continents being biological in origin. Atmospheric N is biologically fixed and then subsequently added to the continents either via collision of marginal marine sediments or incorporation from subduction zone processing, with a total crust N content of 0.5-0.67 PAN. While we do have both of these fluxes in our model, they are only very general. It is well known that the Earth has gone through periods of orogenic activity and periods of quiescence \citep[e.g.,][]{Condie_2013} with variable passive margin extent \citep{Bradley_2008}, and these variations are not captured in our model. 

In all model runs, prior to the evolution of oxygenic photosynthesis the N cycle is marked by mantle degassing and atmospheric N growth, after a period of atmospheric equilibration with oceanic sediments. (Figs. \ref{fig:standardNresv} - \ref{fig:platestyle}). High mantle temperatures, combined with low efficiency export production and N-fixing results in net mantle outgassing and atmospheric growth for the first 1.5 Ga of model output. Biologic N fluxes (Fig. \ref{fig:platephotchange}) are low prior to oxygenic photosynthesis, due to lower overall productivity and burial.  Then, coincident with, and caused by, the appearance of oxygenic photosynthesis, the atmosphere is drawn down, with an increase in all the geologic N reservoirs. Deep ocean sediments increase most quickly, with continental crust and the mantle increasing more slowly. At the GOE, an increase in weathering drives a spike in productivity and further atmospheric draw-down due to enhanced N-fixing. The Lomagundi-Jatuli type event, with high productivity, lasts for $\sim50$ Myr. At about 1.6 Ga, the mantle and atmosphere have equal N budgets, and the mantle continues to increase at the expense of all other reservoirs until the present day. 

While the nominal run reproduces modern N distribution well, there are discrepancies between modeled biologic and geologic fluxes and estimates of these fluxes from the literature (Fig. \ref{fig:standardNresv}). The model somewhat underestimates N-fixing, nitrification, and denitrification. It is possible to explain some of this discrepancy by the lack of continental biologic N-cycling in our model, since  at present continental ecosystems  account for half of global biologic N cycling \citep[e.g.,][]{Gruber_and_Galloway_2008}. Similarly, nominal output for continental N weathering is less than a recent study \citep{Houlton_et_al_2018}. Adding more explicit treatment of continental N cycling would be a welcome addition to this model.  

For the other major geologic fluxes, outgassing and subduction, the EarthN model output is higher than estimates from the literature. This indicates that either subduction of N is not as efficient as we describe, and recycling into the mantle is less, or literature estimates of subduction and outgassing are too low. It is notoriously difficult to estimate N fluxes outgassing at subduction zones and mid-ocean ridges \citep{Fischer_et_al_2002, Elkins_et_al_2006}, due to high background atmospheric N$_2$. In addition, there are very few estimates of N cycling in subduction zones \citep{Fischer_et_al_2002, Elkins_et_al_2006, Mitchell_et_al_2010, Halama_et_al_2014, Mallik_et_al_2018}, and it is possible that this flux is being underestimated in the literature due to the difficulty in analyzing silicate-bound N. 

\subsubsection{Ocean nutrients through time}

Our results yield several important predictions of nutrient content of the ocean through time. First, and unsurprisingly, NH$^+_4$is the dominant bioavailable N species in the ocean, and PO$^{3-}_4$ remains high ($\sim 1-3~\mu$M) before oxygenic photosynthesis. Second, after the appearance of oxygenic photosynthesis and the associated increase in productivity (Fig. \ref{fig:model_forcings}), burial increases. The increase in burial is reflected in the increase in N in ocean sediments  and in all biologic N fluxes (Fig. \ref{fig:standardNresv}). We note here that since we do not include abiotic N-fixing, and instead assume biologic N-fixing could operate throughout the model run, we might be slightly overestimating early Archean N-fixing if biologic fixing did not evolve till 3.2 Ga \citep{Stueken_et_al_2015}. For example, \citet{Navarro_et_al_2001} suggests $2.1\times10^{10}$ mol yr$^{-1}$ could be fixed abiotically by lightning, while the minimum N-fixing we calculate is $2.9\times10^{11}$ mol yr$^{-1}$. At the same time as the increase in N-cycle fluxes, PO$^{3-}_4$ concentrations drop an order of magnitude, again the result of increased productivity and burial. We also note that there is a small oxygen oasis in the shelf ocean box, which is consistent with evidence for localized oxic conditions prior to widespread oxygenation at the GOE \citep{Anbar_et_al_2007} Additionally, nutrient concentrations increase after the GOE, as the result of increased weathering efficiency. 

Third, NH$^+_4$~and NO$^-_3$ ~are at about the same concentration in the Proterozoic. This balance is the result of O$_2$ levels, and so is dependent on our O$_2$ forcing scheme, which sets Proterozoic O$_2$ levels at $1\%$ of modern. There is not agreement on the exact level of O$_2$ in the Proterozoic \citep{Lyons_et_al_2014, Planavsky_et_al_2014, Reinhard_et_al_2016O2, Zhang_et_al_2016}, but generally the maximum estimates are no greater than $10\%$ of modern. Thus, the transition to a NO$^-_3$-rich ocean is predicted to only occur at the NOE, and our modeling does not indicate any sort of N-limitation during the Proterozoic. We do not, however, model a specific increase in productivity due to the evolution of eukaryotes, which might be expected to enhance biologic N-cycling \citep[e.g.,][and references therein]{Zerkle_and_Mikhail_2017}. 

\subsubsection{Different plate tectonic histories}
Different mantle cooling history could have a large effect on the transfer and sequestration of N from the surface into the mantle over time. For example, if mantle temperatures were hotter in the Archean, one might expect both faster mantle convection and less efficient retention of N at subduction zones. Relatedly, if mantle temperature is not the main control on N retention into the mantle at subduction zones, different subduction efficiencies would lead to correspondingly different N histories. 

\begin{figure}
\begin{center}
\includegraphics[width=\textwidth]{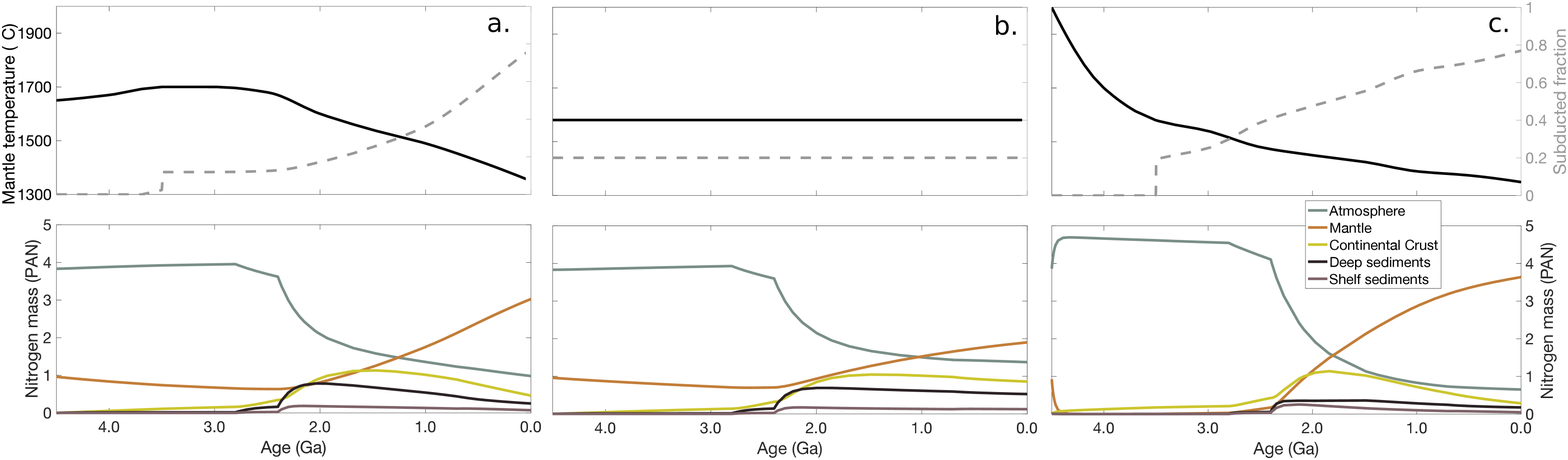}
\caption{Nitrogen reservoirs and subducted fraction for different subduction parameters. a.) nominal run, b.) constant subduction, c.) canonical subduction. The Archean in all cases is characterized by net mantle outgassing, but outgassing is rapid in canonical subduction runs. Net drawdown occurs after oxygenic photosynthesis evolves, and deep sediments and continental crust increases as well.}
\label{fig:platestyle}
\end{center}
\end{figure}

To test these possibilities, we ran the model with three different styles of mantle cooling/plate tectonic transfer of N from the surface to the mantle (Fig. \ref{fig:platestyle}). The first, as described in the nominal run section, is based on \citet{Korenaga_2010}. In the second, mantle temperature, ocean crust production, and subducted crust are all constant, set to the average of each value from  \citet{Sandu_et_al_2011}, which uses a ``canonical'' mantle cooling. In addition, the fraction of subducted N that is transported to the deep mantle is held constant at 0.2. In the third realization, we allow subducted fraction to vary with canonical mantle evolution temperature.  

In all three realizations, prior to oxygenic photosynthesis, there is net mantle outgassing and atmospheric growth. Similarly, after oxygenic photosynthesis and more efficient export production, there is net atmospheric drawdown into geologic reservoirs. The amount of drawdown by subduction with canonical mantle cooling, with either constant or temperature-linked efficiency, is more overall than in the nominal run, up to 4 PAN in the latter. Constant subduction efficiency, however, cannot sequester enough atmospheric N into the mantle to result in a 1 PAN atmosphere at modern. In addition, this run results in more N in the atmosphere than the mantle, contradicting estimates of N distribution on Earth today \citep{Johnson_and_Goldblatt_2015}. Interestingly, the overall pattern is insensitive to mantle cooling history.  

We also explored realizations where the time of plate tectonic initiation and oxygenic photosynthesis were varied (Fig. \ref{fig:platephotchange}). The initiation of plate tectonics does not affect the overall pattern, nor does it greatly effect the distribution of N between various reservoirs. Mantle cooling, and its effect on subduction efficiency, has a larger effect than timing of plate initiation alone. Similarly, only when biologic productivity increases after oxygenic photosynthesis do major changes in N distribution occur.

\subsubsection{Different oxygenic photosynthesis appearance}
\begin{figure}[h]
\begin{center}
\includegraphics[width=\textwidth]{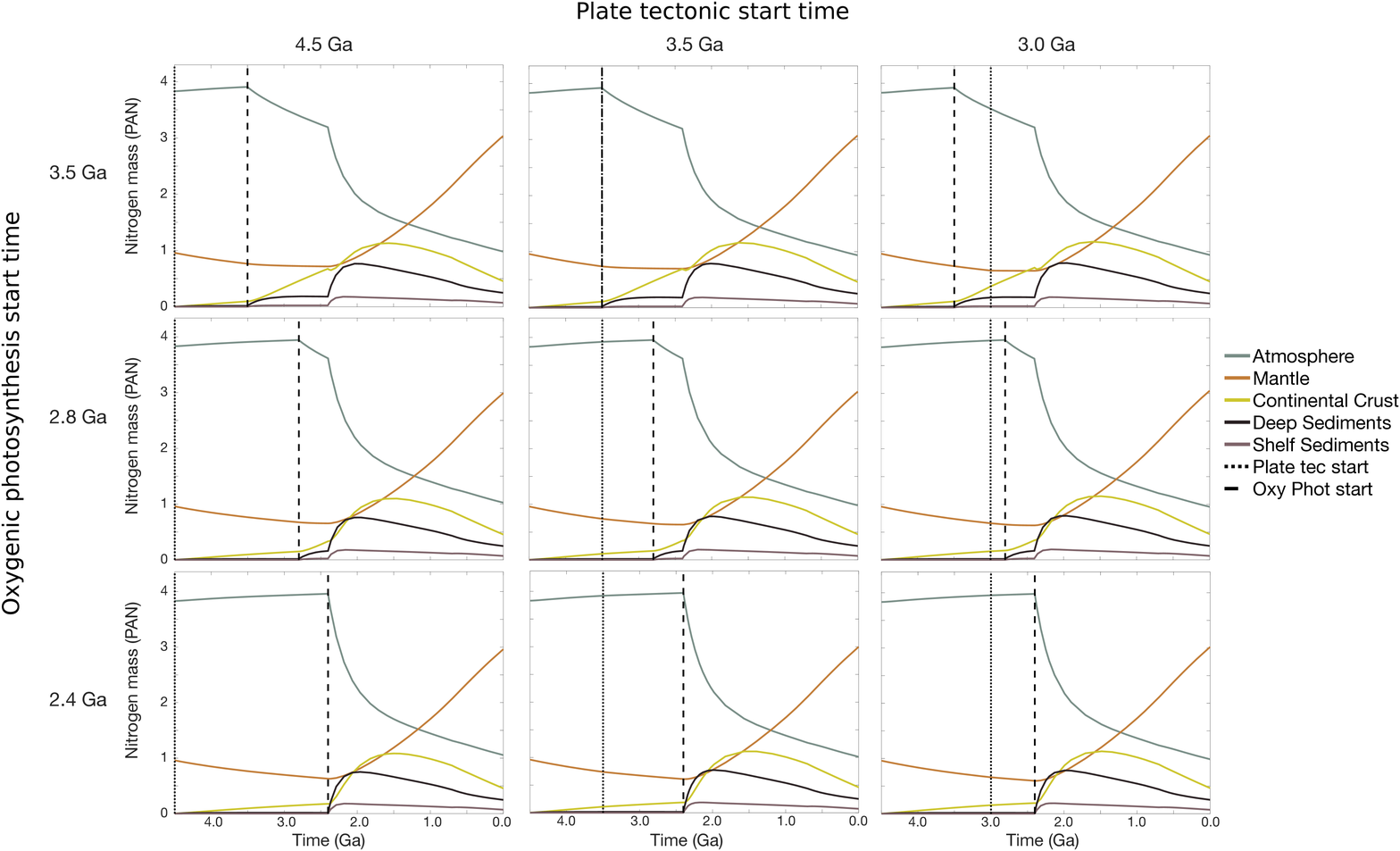}
\caption{Nominal runs varying plate tectonic (4.5, 3.5, 3.0 Ga) and oxygenic photosynthesis starting time (3.5, 2.8, 2.4 Ga). The overall pattern for major N reservoirs is similar for all cases.  Mantle N increases when subduction begins,  atmospheric drawdown of N into ocean sediments occurs when oxygenic photosynthesis evolves, and further drawdown occurs after the GOE.   }
\label{fig:platephotchange}
\end{center}
\end{figure}

Another main ``knob'' on the control panel of the N cycle is how biologic activity processes this in the oceans. As seen in the nominal run, the appearance of oxygenic photosynthesis and the GOE exert a large control over how active N-fixing, nitrification, and denitrification are. To test for any effects of different times of oxygenic photosynthesis evolution (Fig. (Fig. \ref{fig:platephotchange}), we ran the model with standard conditions, but altered when oxygenic photosynthesis evolves: early (3.5 Ga), middle (2.8 Ga, standard), and late (2.4 Ga). These times were chosen to coincide with  early  fossil evidence for photosynthetic life \citep{Hofmann_et_al_1999}, molecular and geochemical evidence for oxygenic photosynthesis by 2.8 Ga \citep[see][]{Buick_2008}, and the GOE at 2.4 Ga \citep{Farquhar_et_al_2000}.

In all runs, when oxygenic photosynthesis evolves, atmospheric N is drawn down. Initially, N is sequestered into deep sediments, then, if plate tectonics is operating, sent into the mantle and continental crust. If oxygenic photosynthesis appears later,  atmospheric N reservoir reaches a slightly higher maximum prior to drawdown. In addition, all else being equal, a later appearance of oxygenic photosynthesis results in a slightly higher pN$_2$ at modern. Overall, the main change in N-history occurs at the GOE, driven by increased weathering, nutrient supply, and enhanced biologic productivity.

\subsection{Atmospheric pN$_2$ comparison with other reconstructions}

As mentioned in the introduction, there is a discord between modern geochecmical data suggesting net ingassing of the atmosphere through time \citep{Busigny_et_al_2011, Nishizawa_et_al_2007, Barry_and_Hilton_2016} and either net outgassing \citep{Som_et_al_2012, Som_et_al_2016} or atmospheric stability since the Archean \citep{Marty_et_al_2013}. Our model strongly suggests dynamic behavior over time, with the N distribution on Earth responding  to changes in biologic and geologic evolution over time. 

\begin{figure}
\begin{center}
\includegraphics[width=\textwidth]{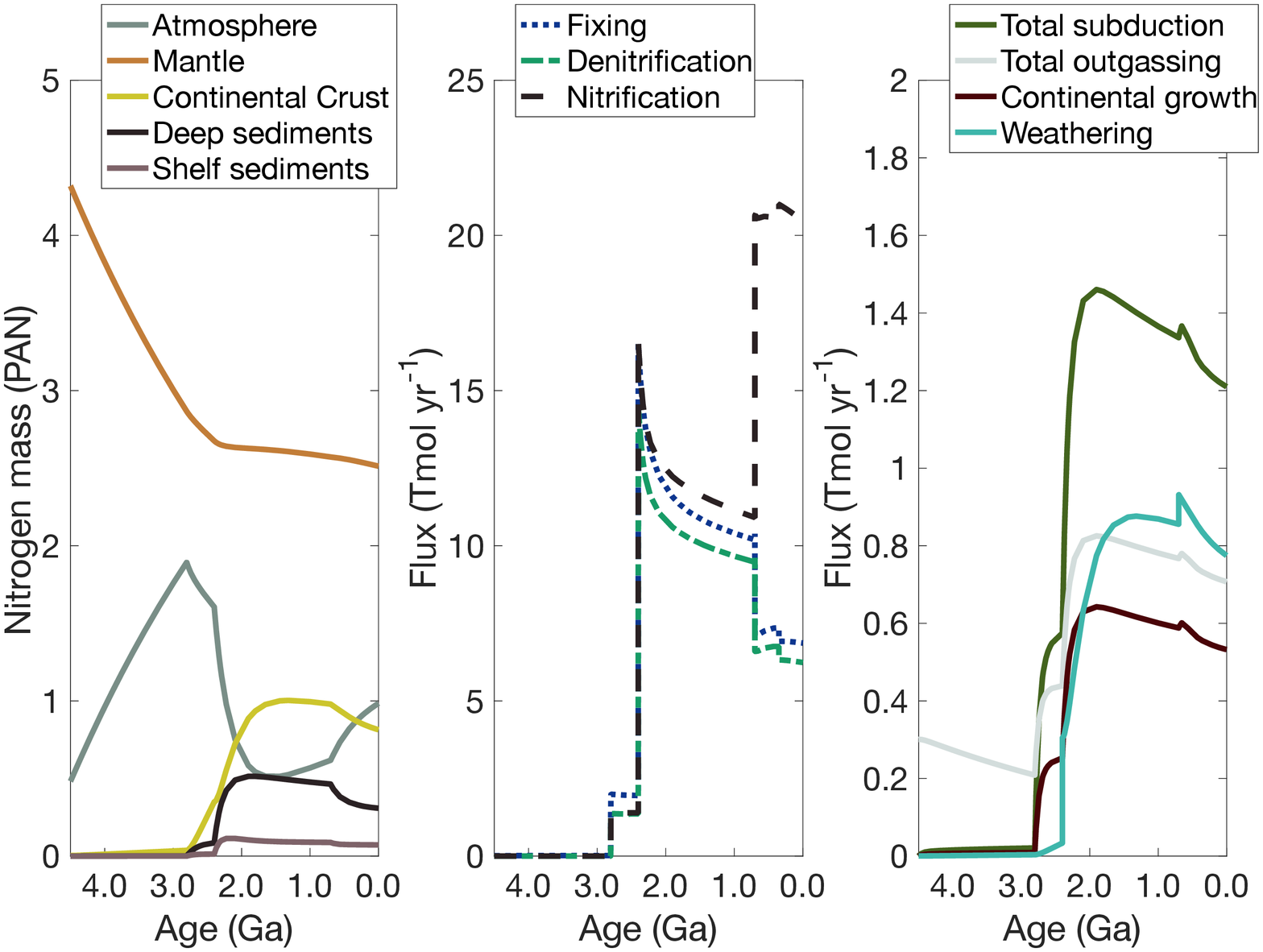}
\caption{Nominal run but with low starting atmospheric mass, $5\%$ starting N in the atmosphere, constant net subduction ($10\%$), and high (50 Sv) hydrothermal circulation.  These conditions lead to a model realization  consistent with previous estimates of a 1 PAN atmosphere at 3.46 Ga \citep{Marty_et_al_2013} and 0.5 PAN at 2.7 Ga \citep{Som_et_al_2012, Som_et_al_2016}.}
\label{fig:lowpN2}
\end{center}
\end{figure}

In order to match constraints for similar or lower atmospheric N mass in the past \citep{Marty_et_al_2013, Som_et_al_2012, Som_et_al_2016}, but still end up with a 1 PAN modern atmosphere, our model has to be tuned to very specific parameters (Fig. \ref{fig:lowpN2}) within an otherwise nominal run. Given our standard total N budget (4.8 PAN), the mantle must start with the majority ($90\%$) of total N, due to net outgassing during the  Hadean and Archean. In addition, the net subduction of N at subduction zones  has to be low and constant ($10\%$), as an increase in net subduction over time results in atmospheric drawdown in all model runs. We also have to increase hydrothermal circulation from 5 to 50 Sv, which limits PO$^{3-}_4$ and in turn limits atmospheric drawdown via N-fixing. The model with these parameters can reproduce the constraints of 1 PAN at 3.46 Ga \citep{Marty_et_al_2013} and a 0.5 PAN atmosphere at 2.7 Ga \citep{Som_et_al_2012, Som_et_al_2016} but still result in a 1 PAN modern atmosphere. Even in this case, there is still a 1.75 PAN atmosphere at 2.8 Ga, when oxygenic photosynthesis evolves. The atmosphere undergoes a dynamic evolution.

Lower atmospheric mass in the past cannot be specifically ruled out by our model output, but such lower mass would present a number of interesting implications. The lack of evidence for large glaciations in the Archean is difficult to reconcile with low atmospheric mass \citep{Goldblatt_et_al_2009}. Similarly, the majority of the Proterozoic, or ``boring billion'', lacks evidence for glaciation, implying warm climate. If the Earth had less than a one PAN atmosphere, there would need to be $10^{-2.2}$ bars of CO$_2$, and even more with less N.

\subsection{Investigating unknown nitrogen distribution: Monte Carlo simulations} 

\begin{figure}[h]
\begin{center}
\includegraphics[width=\textwidth]{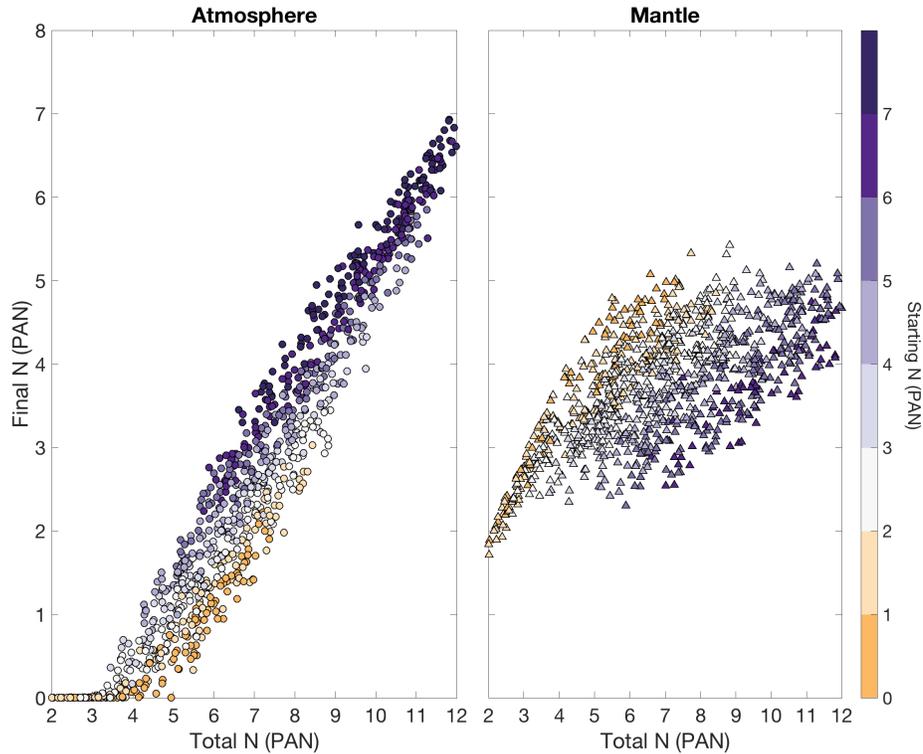}
\caption{Final atmospheric and mantle N vs total N for 1000 Monte Carlo simulations. Note strong correlation between atmospheric N and total N, and that for very low N budgets, the atmospheric mass is small. Runs that results in a 1 PAN atmosphere have 5 PAN total N. The mantle saturates at 3 PAN.}
\label{fig:totalNfinalatm}
\end{center}
\end{figure}

\begin{figure}[h]
\begin{center}
\includegraphics[width=\textwidth]{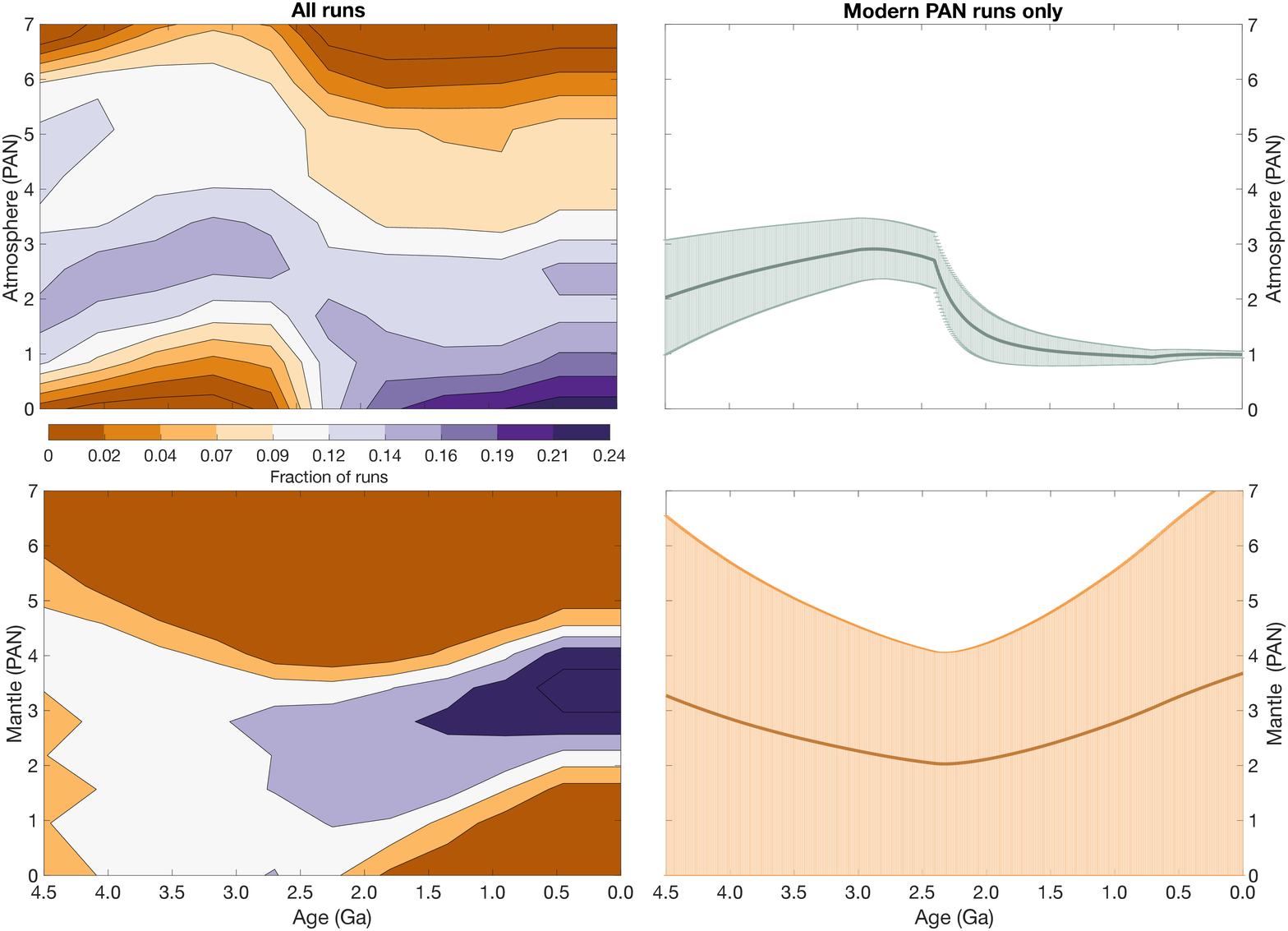}
\caption{Results of Monte Carlo runs for atmosphere and mantle values through time for all runs (a-b) and just runs that result in a 1 PAN present atmosphere (c-d). Panels c-d show the mean and one standard deviation. The mantle tends to evolve towards a final value of between 3-4 over time, while the atmosphere is more variable. Most runs end with a 2 PAN atmosphere at modern and throughout much of the Proterozoic and Archean, with two other nodes at 0.5 and 4 PAN. Runs that result in a 1 PAN atmosphere start with a $\sim2$ PAN atmosphere at 4.5 Ga,, and are drawn down in the late Archean. Mantle N for runs that result in a 1 PAN atmosphere average 4 PAN, but are variable.  }
\label{fig:Ncontour}
\end{center}
\end{figure}

Despite recent interest in geologic and Earth system N cycling \citep{Johnson_and_Goldblatt_2015, Zerkle_and_Mikhail_2017, Johnson_and_Goldblatt_2017}, there is not a consensus on how much N the Earth contains and how it has moved between different reservoirs over time. In addition, and especially in the Hadean and Archean, a number of parameters that could affect the N cycle are not well constrained. These include when oxygenic photosynthesis first appeared, the rate of hydrothermal alteration of ocean crust, deep water upwelling, continental weathering, and the initiation of plate tectonics \citep{van_Hunen_Moyen_2012}. To investigate how changing these poorly constrained parameters may have affected the N cycle over time, we ran  Monte Carlo simulations (n=1000) where a number of parameters were given random values within a prescribed range (Table \ref{tab:MCconditions}, Figs. \ref{fig:totalNfinalatm}-\ref{fig:Ncontour}). 

\begin{table}[htbp]
   \centering
     \caption{Range of values used for Monte Carlo simulartions. PAN is Present Atmospheric Nitrogen, or $4\times10^{18}$ kg}
   \begin{tabular}{@{} l r @{}} 
         \bf{Parameter} & \bf{Range (units)}\\ \hline
         Upwelling & 0.16--16 Sv \\
         Oxygenic photosynthesis start time & 2.4-3 Ga\\
         Plate tectonics start time & 3-4 Ga \\
         Weathering timescale & 50-500 Myr \\
         Hydrothermal flow rate & 0.5-50$\times10^{16}$ L yr$^{-1}$\\
         Total N & 2-12 PAN\\
         Percent starting in atmosphere & 0-100$\%$ \\
      \hline
   \end{tabular}
 
   \label{tab:MCconditions}
\end{table}

Strikingly, our model suggests that the amount of N found in the atmosphere today may in some part be related to the total N in the planet (Fig. \ref{fig:totalNfinalatm}). There is a strong correlation between the total N in the model and the atmospheric N mass after 4.5 Gyr of planetary evolution. Monte Carlo results that have 1 PAN atmosphere at the present day are those with a total N budget of $\sim4-6$ PAN,  similar to independent budget estimates \citep{Johnson_and_Goldblatt_2015}. On Earth, the atmospheric N content directly relates to the total planetary N budget. It is possible that this proxy may work for other terrestrial planets, given evidence for plate tectonics (e.g., linear mountain belts, bimodal topographies) and biologic N cycling (e.g., N$_2$ and O$_2$ coexisting, N$_2$O), atmospheric N could serve as an estimate for total planetary N content. 

The mantle N content also increases with increasing total content, but tends to ``saturate'' at 4.5 PAN (Fig. \ref{fig:Ncontour}). This is because net subduction and outgassing at mid-ocean ridges tend to balance each other, while N-fixing is limited by PO$^{3-}_4$ availability. That is, the rate at which organisms can fix N is not enough to outpace overall outgassing, thus at higher total N budgets, N accumulates in the atmosphere. For very small total N budgets less than 3 PAN, N-fixing is efficient enough to draw down the atmosphere almost completely into the mantle. 

In addition, the EarthN  model does not thoroughly parameterize  mantle evolution over time. Thus, we predict that model output suggesting mantle saturation at 4.5 is a minimum estimate of the true N content of the mantle.. Previous work has estimated that the mantle has an enormous capacity for N, and could possibly contain many 10s of PAN \citep{Li_et_al_2013, Smith_and_Kopylova_2014, Johnson_and_Goldblatt_2015}. 

Monte Carlo realizations that result in a 1 PAN present atmosphere show the same overall pattern as the nominal run (Fig. \ref{fig:Ncontour}). On average,  the atmosphere starts out with greater than modern mass, then decreases, again likely due to the evolution of oxygenic photosynthesis, in the late Archean. The atmosphere remains at approximately 1 PAN throughout the Proterozoic, with a small increase at the NOE, due to enhanced denitrification at this time. Even when we allow for a random total N and a random amount of N starting in the atmosphere, the overall trend in the atmospheric evolution is to be drawn down from higher Archean values towards the modern. In contrast, the mantle stays consistent on average, at its saturation of about 4 PAN. The standard deviation shown in Fig. \ref{fig:Ncontour}-d indicates there are many mantle N paths which are consistent with the EarthN model evolving a 1 PAN present atmosphere.

As previously mentioned, the temperature history of the mantle, and behavior of N in different geothermal regimes exerts a strong control on the evolution of N in the Earth system over time. This behavior also depends on how much temperature actually controls N volatilization at subduction zones. Studies have shown that redox \citep{Li_et_al_2013}, temperature/pressure \citep{Li_and_Keppler_2014}, distribution of N between fluid and melt \citep{Mallik_et_al_2018}, and pH \citep{Mikhail_and_Sverjensky_2014} also exert control over N speciation and retention in the mantle. The interaction of these different factors in geologic evolution could have had a large effect on N cycling over Earth history \citep{Mikhail_and_Howell_2016}. We predict that redox evolution is likely to have the largest effect on N processing and storage in the mantle. The upper mantle and mantle lithosphere have, on average, been at their current redox state since 3.5 Ga \citep{Canil_2002}. Arc basalts, however, are more oxidized than  MORBs with otherwise similar chemical composition \citep{Kelley_and_Cottrell_2012, Brounce_et_al_2014}. In addition, the deeper mantle is more reducing than the upper mantle \citep{Frost_and_McCammon_2008}. 

It follows, then, that N is more soluble in the reduced lower mantle than the more oxidized upper mantle and subduction zone mantle wedge. As the mantle has become more oxidized through time, N recycling to the atmosphere would be favored. If the oxidation state of subduction zone mantle has similarly increased through time, this would enhance recycling of N to the surface \citep[e.g.,][]{Mikhail_and_Sverjensky_2014}. Thus, there are competing features controlling N recycling into the mantle: decreasing temperature and increasing $f$O$_2$. Given observations of net N retention at modern subduction zones \citep{Li_et_al_2007,Mitchell_et_al_2010, Halama_et_al_2014}, we hypothesize that mantle wedge temperature is the dominant control of N recycling efficiency, at least on the modern Earth. How this balance of redox and temperature has controlled N, and other volatile, recycling over Earth history has important implications for the evolution of the surface and interior of the planet. 

The above discussion highlights a broader point regarding model construction. Herein, we have constructed an Earth system N cycle model, and presented a nominal run based on plausible assumptions about the Earth through time. The results presented, however, should not be taken as gospel, dogma, or actuality. There are a number of fluxes and factors in the Earth which could affect results of the EarthN model. In addition to mantle chemistry, mantle capacity for N is enormous \citep{Li_et_al_2013}, and the great potential size of this reservoir could have major influence over N cycling during Earth history. The lack of continental ecosystems, and simple treatment of hydrothermal activity in oceanic crust in this model could be important parameters to investigate. The addition of isotopes to the model would allow for predictions that could be tested in the rock record. We envision future studies to explore this wider parameter space, both for specific intervals in time and for grand trends over Earth history.

\section{Conclusions}

We have constructed an Earth system N model, EarthN, that includes biologic and geologic fluxes to predict the distribution of N in the major reservoirs of the Earth through time. In addition to linking the N cycle to PO$^{3-}_4$ availability, the model is driven by changing O$_2$ abundance and mantle cooling with plate tectonics. Model output is consistent with movement of N between the three major reservoirs (atmosphere, mantle, continental crust) in significant amounts over Earth history. 

In all model runs, the early part of Earth history, from 4.5-2.8 Ga, is characterized by net mantle outgassing and atmospheric growth. This early history is due to high mantle temperatures and inefficient export production. After the evolution of oxygenic photosynthesis, atmospheric N is immediately drawn down and sequestered in sediments due to increased export production.  At the Great Oxidation Event, increased weathering and nutrient delivery enhances export production, which in turn enhances atmospheric draw-down via N-fixation. Mantle cooling over time, with associated increase in efficiency of N subduction, facilitates biologically fixed N to be sequestered into geologic reservoirs over time.

One of the strongest controls on the atmospheric mass of N through time, and especially the modern mass of the atmosphere, is the total N in the Bulk Silicate Earth. Monte Carlo simulations that vary a number of parameters (deep water upwelling, hydrothermal circulation, oxygenic photosynthesis appearance, weathering timescale, total N and distribution) that result in a 1 PAN atmosphere after 4.5 Ga of model evolution are most consistent with a total BSE N budget of $\sim4-6$ PAN. The mantle tends to saturate at 4-4.75 PAN. The mantle is the dominant N carrier for total N budgets below 6-7 PAN, while the atmosphere is dominant at higher values.  

The EarthN model shows that the distribution of N in the Earth system through time could have varied significantly. Nominal model runs result in net atmospheric drawdown over time, which is consistent with geochemical proxies. There are a number of controls on  N history,  including appearance of oxygenic photosynthesis, mantle cooling, and N  in subduction zones. We anticipate further work in this area to focus on how temperature and redox control N at subduction zones. Equally, the cycling of N in the mantle over time is poorly known but crucially important. There is potential for not only investigating Earth history, but exploration of Venus, Mars, and potential exoplanetary targets in the future. 

%  ACKNOWLEDGMENTS

\acknowledgments
The authors would like to acknowledge Katja Fennel for sharing code. We also acknowledge helpful discussions concerning model development with Rameses D'Souza, Arlan Dirkson, and Christiaan Laureijs at the University of Victoria. Ananya Mallik and one anonymous reviewer are thanked for useful reviews, as is Cyn-Ty Lee for editorial duties. 

Supporting model code and output data can be found as a supplemental file with this manuscript. In addition, code will be available at the corresponding author's webpage: \href{http://www.benwjohnson.com}{www.benwjohnson.com}. 

BWJ is currently supported by NSF (EAR - 1725784) and was previously supported by NSERC Discovery grant to CZG. CZG is supported by an NSERC Discovery grant. 
\appendix

\section{Differential equations}
\label{apdx:Diff_eqns}

Based on the above model decsription, we write a series of differential equations to solve for model species in  boxes. 
\begin{linenomath*}

\textbf{Atmosphere}
\begin{eqnarray}
\frac{dR_j^\text{atm}}{dt} & = & \sum_{i*}F_\text{as,j}^{atm-i*} + F_\text{ogarc,j} 
\end{eqnarray}
for $j=$\{N$_2$,$^{40}$Ar,$^{36}$Ar\} and $i*$ includes air-sea flux from all shallow ocean boxes (\{low, high, shelf\}). 

 \textbf{Low- and high-latitude shallow ocean} 
\begin{eqnarray}
\frac{dR_\text{40Ar}^i}{dt} &=& F_\text{mix,40Ar}^i +  F_\text{rd}^i - F_\text{as,40Ar}^{\text{atm}-i}\\ 
\frac{dR_\text{36Ar}^i}{dt} &=& F_\text{mix,36Ar}^i - F_\text{as,36Ar}^{\text{atm}-i} \\ 
\frac{dR_\text{40K}^i}{dt}  &=& F_\text{mix,40K}^i  -  F_\text{rd}^i \\ 
\frac{dR_\text{K}^i}{dt}  &=& F_\text{mix,K}^i   \\ 
\frac{dR_\text{PO4}^i}{dt}  &=& F_\text{mix,PO4}^i  - F_\text{ExT}^i \\ 
\frac{dR_\text{NO3}^i}{dt}  &=& F_\text{mix,NO3}^i  - F_\text{out,NO3}^i + F_\text{nit}^i - F_\text{den}^i \\ 
\frac{dR_\text{NH4}^i}{dt}  &=& F_\text{mix,NH4}^i  - F_\text{out,NH4}^i - F_\text{nit}^i \\ 
\frac{dR_\text{N2}^i}{dt}  &=& F_\text{mix,N2}^i  - F_\text{out,N2}^i + F_\text{den}^i 
\end{eqnarray}
for $i=$\{low, high\}.

 \textbf{Shelf ocean}
\begin{eqnarray}
\frac{dR_\text{36Ar}^\text{shelf}}{dt} &=& F_\text{mix,36Ar}^\text{shelf}  - F_\text{as,36Ar}^\text{atm-shelf} \\ 
\frac{dR_\text{40Ar}^\text{shelf}}{dt} &=&  F_\text{rd}^\text{shelf} + F_\text{mix,40Ar}^\text{shelf} - F_\text{as,40Ar}^\text{atm-shelf}  \\ 
\frac{dR_\text{40K}^\text{shelf}}{dt}  &=& F_\text{mix,40K}^\text{shelf}  -  F_\text{rd}^i + F_\text{w,40K} \\ 
\frac{dR_\text{K}^\text{shelf}}{dt}  &=& F_\text{mix,K}^i + F_\text{w,K}  \\ 
\frac{dR_\text{PO4}^\text{shelf}}{dt}  &=& F_\text{mix,PO4}^\text{shelf}  - F_\text{ExT}^\text{shelf} + F_\text{remin,PO4}^\text{shelf} + F_\text{w,PO4} - F^\text{shelf}_\text{NO3}\\ 
\frac{dR_\text{NO3}^\text{shelf}}{dt}  &=& F_\text{mix,NO3}^\text{shelf}  - F_\text{out,NO3}^\text{shelf} + F_\text{nit}^\text{shelf} - F_\text{den}^\text{shelf} + F_\text{remin,NO3}^\text{shelf}  - F^\text{shelf}_\text{NO3} \\ 
\frac{dR_\text{NH4}^\text{shelf}}{dt}  &=& F_\text{mix,NH4}^\text{shelf}  - F_\text{out,NH4}^\text{shelf} - F_\text{nit}^\text{shelf} + F_\text{remin,NH4}^\text{shelf} + F_\text{w,NH4} - F_\text{seddif}^\text{shelf}\\
\frac{dR_\text{N2}^\text{shelf}}{dt}  &=& F_\text{mix,N2}^\text{shelf} - F_\text{out,N2}^\text{shelf} + F_\text{den}^\text{shelf} 
\end{eqnarray}

  \textbf{Deep ocean}
\begin{eqnarray}
\frac{dR_\text{40Ar}^\text{deep}}{dt} &=& F_\text{mix,40Ar}^\text{deep} +  F_\text{rd}^\text{deep} - F_\text{hydro,40Ar}^{\text{deep}} +  F_\text{ogmor,40Ar}\\ 
\frac{dR_\text{36Ar}^\text{deep}}{dt} &=& F_\text{mix,36Ar}^\text{deep}  - F_\text{hydro,36Ar}^{\text{deep}} +  F_\text{ogmor,36Ar}\\ 
\frac{dR_\text{40K}^\text{deep}}{dt}  &=& F_\text{mix,40K}^\text{deep}  -  F_\text{rd}^\text{deep} - F_\text{hydro,40K} +  F_\text{ogmor,40K} \\ 
\frac{dR_\text{K}^\text{deep}}{dt}  &=& F_\text{mix,K}^\text{deep} - F_\text{hydro,K} +  F_\text{ogmor,K}\\ 
\frac{dR_\text{PO4}^\text{deep}}{dt}  &=& F_\text{mix,PO4}^\text{deep}  + F_\text{remin,PO4}^\text{deep} - F_\text{hydro,PO4}  + F_\text{ogmor,PO4}\\ 
\frac{dR_\text{NO3}^\text{deep}}{dt}  &=& F_\text{mix,NO3}^\text{deep} + F_\text{remin,NO3}^\text{deep} + F_\text{nit}^\text{deep} - F_\text{den}^\text{deep} - F_\text{hydro,NO3}^\text{deep}  \\ 
\frac{dR_\text{NH4}^\text{deep}}{dt}  &=& F_\text{mix,NH4}^\text{deep} - F_\text{nit}^\text{deep} + F_\text{remin,NH4}^\text{deep}  - F_\text{seddif}^\text{deep} - F_\text{hydro,NH4}^\text{deep}\\ 
\frac{dR_\text{N2}^\text{deep}}{dt}  &=& F_\text{mix,N2}^\text{deep}  + F_\text{den}^\text{deep} + F_\text{ogmor,N}
\end{eqnarray}

  \textbf{Sediments}
\begin{eqnarray}
\frac{dR_\text{NO3}^i}{dt}  &=& F_\text{nit}^i - F_\text{den}^i + F_\text{seddif,NO3}^{iii}  \\ 
\frac{dR_\text{NH4}^i}{dt}  &=&   F_\text{remin,NH4}^i - F_\text{nit}^i + F_\text{seddif,NH4}^{iii}  \\ 
\frac{dR_\text{PO4}^i}{dt}  &=&  F_\text{remin,PO4}^i + F_\text{seddif}^{iii}  \\ 
\frac{dR_\text{N}^{ii}}{dt}  &=&  F_\text{bur,NH4}^i  - F_\text{sub,NH4}^{ii} \\ 
\frac{dR_\text{PO4}^{ii}}{dt}  &=&  F_\text{bur,PO4}^i - F_\text{sub,PO4}^{ii} 
\end{eqnarray}
for $i=$\{reactive shelf sediments, reactive deep sediments\}, $ii=$\{shelf sediments, deep sediments\}, and $iii=$\{shelf ocean,  deep ocean\}.

 \textbf{Ocean crust}
\begin{eqnarray}
\frac{dR_\text{40Ar}^i}{dt}& =  &F_\text{rd} - F_\text{sub,40Ar}^i\\ 
\frac{dR_\text{40K}^i}{dt}  &= & -  F_\text{rd} - F_\text{sub,40K}^i \\ 
\frac{dR_\text{K}^i}{dt}  &=& F_\text{hydro,K} - F_\text{sub,K}^i \\ 
\frac{dR_\text{N}^i}{dt}  &= & F_\text{hydro,N} - F_\text{sub,NH4}^i \\ 
\frac{dR_\text{PO4}^i}{dt}  &=&  F_\text{hydro,PO4} - F_\text{sub,PO4}^i 
\end{eqnarray}
for $i=$\{ocean crust\}.

 \textbf{Continental crust}
\begin{eqnarray}
\frac{dR_\text{40Ar}^i}{dt} &=&  F_\text{rd}^i - F_\text{w,40Ar}\\ 
\frac{dR_\text{40K}^i}{dt}  &=&  -  F_\text{rd}^i - F_\text{w,40K} \\ 
\frac{dR_j^i}{dt}  &=&  F_\text{cg,j}  - F_\text{w,j}  
\end{eqnarray}
for $i=$\{continental crust\} and $j$=\{K,~NH$^+_4$,~PO$^{3-}_4$\}.

 \textbf{Mantle}
\begin{eqnarray}
\frac{dR_\text{40Ar}^i}{dt} &=&  F_\text{rd}^i - F_\text{ogm,40Ar} + F_\text{subnet,40Ar}\\
\frac{dR_\text{40K}^i}{dt}  &=&  -  F_\text{rd}^i - F_\text{ogm,40K} +F_\text{subnet,40K}\\
\frac{dR_j^i}{dt}  &=& - F_{\text{ogmor},j} + F_{\text{subnet},j}
\end{eqnarray}
for $i=$\{mantle\} and $j=\{^{36}\text{Ar, N, PO}^{3-}_4, \text{and K}\}$. 
\end{linenomath*}

\section{Calculating initial  N distribution}
\label{apdx:Initial_N}
\begin{figure}[h]
\begin{center}
\includegraphics[width=\textwidth]{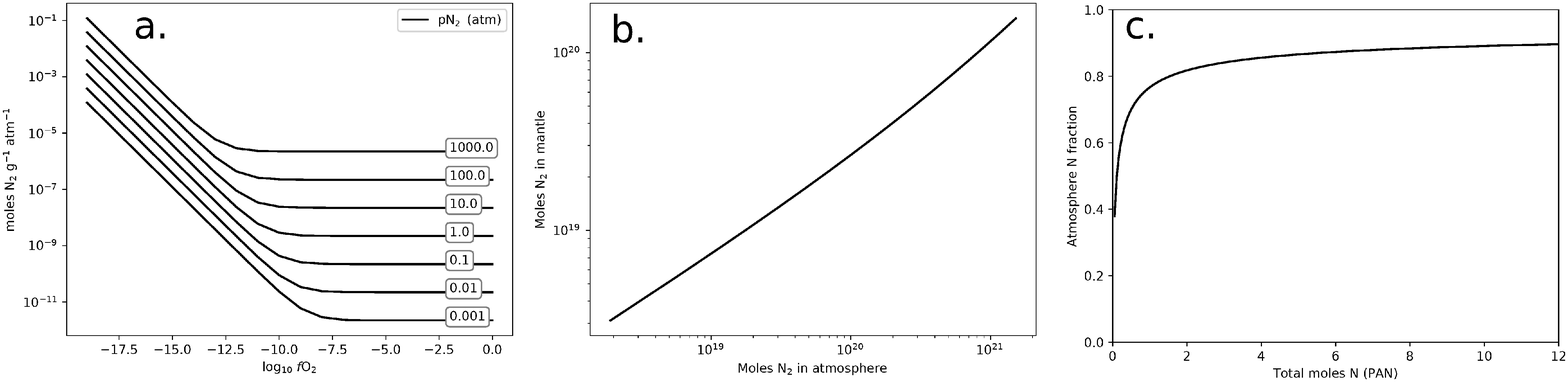}
\caption{Nitrogen solubility in basaltic magma used to calculate starting atmospheric N. a.) Basaltic magma N$_2$ content as a function of oxygen fugacity. Lines are for different pN$_2$, with pressure in atmospheres shown in boxes. Modified after \citet{Libourel_et_al_2003}. b.) Total mantle and atmospheric N based on the solubility experiments of \citet{Libourel_et_al_2003} at the end of core formation. c.) Atmospheric fraction of N as a function of total N (PAN) for an atmosphere in equilibrium with the mantle at the end of the magma ocean phase. Oxygen fugacity is IW - 2. }
\label{fig:Initial_calc}
\end{center}
\end{figure}

We use the following equation from \citet{Libourel_et_al_2003} to calculate the starting amount of N in the atmosphere and mantle for the nominal run: 

\begin{equation} 
N_{2\text{, magma}} = 2.21\times10^{-9} \text{pN}_2+fO_2^{-0.75} 2.13\times10^{-17} \text{pN}_2^{0.5}
\end{equation}

\noindent where pN$_2$ is in atmospheres and $N_{2\text{, magma}}$ is in mol g$^{-1}$ atm$^{-1}$. We chose an $f$O$_2$ of IW - 2 ($f$O$_2$ = $10^{-11.4}$), where IW is the iron w\"{u}stite buffer, which is the expected oxygen fugacity of the magma ocean immediately after core formation \citep{Wood_et_al_2006} (Fig. \ref{fig:Initial_calc}). 

Then, for a variety of pN$_2$ values, we calculate a magma N$_2$ concentration  at $f$O$_2$ = IW-2, and multiply this concentration by the mass of the mantle. This assumes the whole mantle equilibrated during the atmosphere during the magma ocean phase.

Finally, we describe the fraction of the total N budget that is in the atmosphere based on the above solubility  calculations. This figure ultimately guided our choice for total N budget and starting atmospheric N mass in the nominal run. We chose 4.5 PAN as the total budget, which results in $82\%$ of the total N starting in the atmosphere. The total N budget was chosen so that the nominal run resulted in a 1 PAN modern atmosphere at the end of the model run.

\section{Argon and potassium model performance checks}
\label{apdx:Ar_K}
\begin{figure}[h]
\begin{center}
\includegraphics[width=\textwidth]{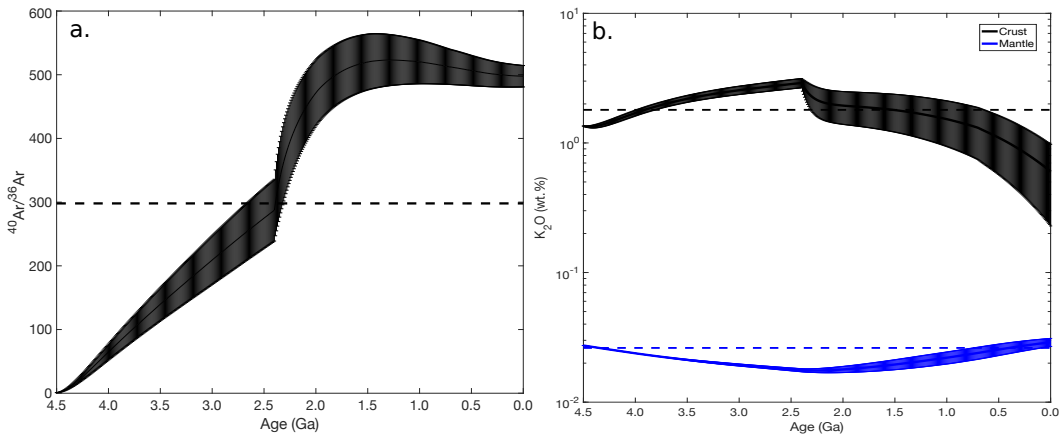}
\caption{Atmospheric$^{40}$Ar/$^{36}Ar$ and  K concentrations (wt. \%K$_2$O) for continental crust and mantle from Monte Carlo runs. The mean and one standard deviation are shown, and modern  values are given as dashed line \citep{Rudnick_and_Gao_2014, Arevalo_et_al_2013}.}
\label{fig:Physical_check}
\end{center}
\end{figure}

We use argon concentration in the atmosphere as a check on the performance of  degassing and air-sea gas exchange in the model. The model overestimates the modern day ratio of $^{40}$Ar/$^{36}Ar$ in the atmosphere by about 1.5 fold (Fig. \ref{fig:Physical_check}). It is possible that this slight overestimate is due in part to the model underestimating K concentration in the continental crust (Fig. \ref{fig:Physical_check}). Higher K-content in the crust would lead to more$^{40}$Ar in the crust, through storage after radioactive decay of$^{40}$K, and would then lower the atmospheric Ar-ratio. 

While the model output reproduces the K content of the mantle well, it underestimates the continental crust concentration. Weathering is simply proportional to concentration in the model, and does not take into account either differences in weathering due to biologic activity or continental growth over time. Given different amounts of crustal growth over time \citep[e.g.,][]{Dhuime_et_al_2012}, the continents would evolve in their capacity to store K over time. In addition, we do not consider the effects of continental lithospheric roots or cratonic mass, which could serve to store K for long periods during Earth history. Future iterations of the EarthN model should incorporate crustal growth scenarios.

\end{document}